\RequirePackage{fix-cm}
\documentclass[smallcondensed]{svjour3}
\smartqed
\usepackage[T1]{fontenc}
\usepackage{times,graphicx,amsmath,amssymb,amsfonts,enumerate,cite,float,marvosym,geometry,needspace}
\usepackage[colorlinks,bookmarksopen,bookmarksnumbered,citecolor=blue,linkcolor=blue,urlcolor=blue]{hyperref}
\graphicspath{{figures/}}
\allowdisplaybreaks[2]
\begin{document}
\begin{sloppypar}

\title{A generalized long-wave limit method with spectral perturbations
}

\titlerunning{A generalized long-wave limit method with spectral perturbations}        

\author{Tianwei Qiu \and Zhen Wang}

\authorrunning{T. Qiu et al.} 

\institute{T. Qiu \at
              School of Mathematical Sciences, Beihang University, Beijing 100191, China \\
           \and
           Z. Wang \Letter \at
              School of Mathematical Sciences, Beihang University, Beijing 100191, China \\
              \email{wangzmath@163.com}           
}

\date{}

\setlength{\headerboxheight}{0pt}
\renewcommand{\makeheadbox}{}
\maketitle

\begin{abstract}
We develop a generalized long-wave limit method for constructing higher-order (degenerate) lump solutions of $(2+1)$-dimensional integrable systems. Starting from the multi-soliton tau function, we introduce wavenumber rescalings, spectral perturbations, and higher-order phase modulations in a single small parameter, and choose their coefficients so that the lower-order Taylor coefficients vanish. Applying this framework to the Kadomtsev--Petviashvili I equation, we explicitly derive second-order two- and three-lump solutions as well as third-order three-, four-, five-, and six-lump solutions. The long-time asymptotics of the peak positions and amplitudes describe anomalous scattering of these degenerate lump solutions. Extensions to hybrid higher-order lump solutions associated with multiple distinct spectral parameters are also discussed.
\end{abstract}

\section{Introduction}
\label{intro}

The Kadomtsev--Petviashvili (KP) equation describes weakly nonlinear, weakly dispersive waves with slow transverse variations \cite{kp1970,kpwater}. Its standard form is
\begin{equation}\label{kp1}
(u_t+6uu_x+u_{xxx})_x+3\sigma^2u_{yy}=0, \quad \sigma^2=\pm1.
\end{equation}
In the context of shallow-water waves, $\sigma^2=-1$ corresponds to the KP-I equation for regimes with strong surface tension, whereas $\sigma^2=1$ yields the gravity-dominated KP-II equation. Line solitons are transversely stable for KP-II \cite{KP2stable}, and their interactions have been observed in shallow coastal waters \cite{KP2ocean}. For KP-I, line solitons are transversely unstable and can break up into two-dimensional localized structures known as lumps under transverse perturbations \cite{raodong}.

Lump solutions are nonsingular rational solitons that decay algebraically in all spatial directions. Satsuma and Ablowitz obtained explicit KP-I lump solutions by taking a long-wave limit of multi-soliton solutions \cite{ablo-lump}. Mathematically, KP-I lumps are closely connected to fully localized solitary waves of the three-dimensional gravity--capillary water-wave problem \cite{lump-Euler}, and the traveling lump has been rigorously proved to be both unique and orbitally stable \cite{lumpunique,lumpstable}. Gravity--capillary lumps have been generated in water-wave tanks using a moving pressure source \cite{lumpexperiment,lump-deepwater1}. Related localized excitations have also been studied in helium films \cite{superfluid}, Bose--Einstein condensates \cite{GPlump,BEClump}, and nonlinear optical systems \cite{KodamaPRL}. The propagation and elastic collisions of optical lumps have been observed in photorefractive crystals \cite{opticlump-PRL}.

Lump solutions can be constructed by the inverse scattering transform (IST) \cite{KP-IST,KP-rh}, Hirota's bilinear method \cite{longwave,quad-mwx}, and the binary Darboux transformation \cite{PLABDT,hjsPRE}. Standard lump solutions, associated with simple discrete eigenvalues in the IST, consist of individual lumps traveling at distinct asymptotic velocities. They undergo normal (elastic) scattering, emerging from mutual collisions with their shapes and velocities intact and without any phase shift \cite{ablo-lump}. Higher-order poles of the meromorphic eigenfunction give rise to multi-pole lump solutions in the IST \cite{multipole,DSmulti,our-CSF}. In a degenerate group associated with a common spectral parameter, the lumps share the same asymptotic velocity, separate sublinearly in time, and exhibit anomalous scattering \cite{JETP1993}. For certain classes of higher-order lump solutions, the asymptotic lump patterns are determined by the zeros of the Yablonskii--Vorob'ev polynomials \cite{yjkkp,integer-partition,Ling-KP}. Degenerate lump solutions have also been constructed by reductions of the KP hierarchy \cite{dmitry1,dmitry2,dyn-KP,class-KP} and by the improved long-wave limit method \cite{anomalous-kp,ours-DJKM,qiu-mkp}. 

IST and Darboux constructions use the Lax pair, while KP hierarchy reductions use determinant representations. The long-wave limit method starts from the explicit multi-soliton tau function and reduces the limiting calculation to algebraic conditions on its parameters. Two limiting procedures have been used to construct degenerate lump solutions within the bilinear framework. One first takes the long-wave limit to obtain a standard lump solution and then coalesces its spectral parameters with suitable phase adjustments \cite{JETP1993,multipole}. The other is the improved long-wave limit method, which allows the wavenumbers to vanish with distinct proportionality factors at a common spectral parameter; this mechanism was originally introduced to generate higher-order rogue waves \cite{ZhangRW} and later adapted to construct specific higher-order lumps in the KP-I and modified KP-I equations \cite{anomalous-kp,qiu-mkp}. In this paper, we unify and generalize these mechanisms by combining wavenumber rescalings, spectral perturbations, and higher-order phase modulations within a single long-wave limit of the explicit multi-soliton solution.

Starting from the $2M$-soliton tau function, we obtain second-order two- and three-lump solutions for $M=2$ and third-order three-, four-, five-, and six-lump solutions for $M=3$. The construction recovers the triangular three- and six-lump solutions in \cite{anomalous-kp} and yields the other degenerate lump solutions through spectral perturbations. We derive the long-time asymptotics of the lump positions and amplitudes for the solutions constructed here. We also describe the degeneration of lump chains into localized higher-order lump solutions.

The remainder of this paper is organized as follows. Section~\ref{sec:2} briefly reviews the classical long-wave limit and its connection to the IST, and then formulates the generalized long-wave limit method with spectral perturbations. Section~\ref{sec:3} presents the construction and asymptotic analysis of second-order lump solutions, together with the degeneration of two-lump chains. Section~\ref{sec:4} derives and analyzes the four families of third-order lump solutions. Finally, Section~\ref{sec:5} summarizes our findings and outlines several extensions.

\section{Preliminaries} \label{sec:2}

We first recall the classical long-wave limit and its relation to the IST, and then introduce the generalized long-wave limit with spectral perturbations.

\subsection{Soliton solutions and the classical long-wave limit method} \label{sec:21}

Hirota's bilinear method \cite{hirotabilinear} gives the $N$-soliton solution of \eqref{kp1}, with $\sigma^2=-1$, in the form
\[
u = 2(\ln f_N)_{xx}.
\]
The tau function $f_N$ satisfies Hirota's bilinear equation
\begin{equation}\label{kp-bilinear}
 (D_xD_t+D_x^4-3D_y^2)f_N\cdot f_N=0,
\end{equation}
where $D_x$, $D_y$ and $D_t$ are Hirota's bilinear differential operators. The $N$-soliton tau function is obtained by summing over $\chi_j\in\{0,1\}$, $j=1,\ldots,N$:
\begin{equation*}
\begin{aligned}
&f_N=\sum_{\chi_1,\ldots,\chi_N=0,1}\exp\left(\sum_{j=1}^N\chi_j\eta_j+\sum_{1\leq j<l\leq N}\chi_j\chi_l\ln a_{jl}\right),\\
&\eta_j=k_jx+p_jy+\left(\frac{3p_j^2}{k_j}-k_j^3\right)t+\eta_{0,j},\\
&a_{jl}=\frac{k_j^2k_l^2(k_j-k_l)^2+(k_jp_l-k_lp_j)^2}{k_j^2k_l^2(k_j+k_l)^2+(k_jp_l-k_lp_j)^2}.
\end{aligned}
\end{equation*}

To take the long-wave limit of the $2M$-soliton tau function, we use the parameterization
\begin{equation} \label{assume}
\begin{aligned}
    &k_j = \beta_j\varepsilon, \quad p_j = \lambda_j k_j, \quad \eta_{0,j} = \ln C_{j0} + \sum_{s=1}^\infty C_{js}\varepsilon^s, \\
    &\beta_{j+M} = \beta_j^*,\quad \lambda_{j+M} = \lambda_j^*, \quad \eta_{0,j+M}=\eta_{0,j}^*, \quad j = 1,2,\dots,M,
\end{aligned}
\end{equation}
where $\varepsilon\in\mathbb{R}$ and the asterisk denotes complex conjugation. In the classical long-wave limit method, one further sets
\[
    \beta_j=1, \quad\eta_{0,j}=\ln{(-1)}+C_{j1}\varepsilon,\quad  C_{j+M,1}=C_{j1}^*,\quad j=1,2,\dots,M,
\]
and takes the limit $\varepsilon\to0$. This yields the asymptotic expansion
\[
    f_{2M}=\tau_{2M}\,\varepsilon^{2M}+O(\varepsilon^{2M+1}).
\]
The limit gives the standard $M$-lump solution
\begin{equation} \label{normal}
    u=2(\ln{\tau_{2M}})_{xx}.
\end{equation}

Throughout this work, the position of an asymptotically separated fundamental lump is defined as the location of its positive local maximum, and the value of $u$ at this point is referred to as its amplitude. For $M=1$, the fundamental lump exhibits the following properties.

\begin{proposition} \label{1lumpsolu}
For the one-lump solution with $\lambda_1\in \mathbb{C}^+$ and $C_{11}=0$, the following properties hold.
\begin{enumerate}
\item[(i)] The solution is a traveling wave moving with constant velocity $\mathbf{v} = \left( 3|\lambda_1|^2,\ -6\mathrm{Re}(\lambda_1) \right)$. The map $\lambda_1 \mapsto \mathbf{v}$ is injective. 
\item[(ii)] For each fixed $t$, the solution is $O\left((x^2+y^2)^{-1}\right)$ as $x^2+y^2\to\infty$.
\item[(iii)] The amplitude of the lump is $4[\mathrm{Im}(\lambda_1)]^2$.
\end{enumerate}
\end{proposition}

For pairwise distinct spectral parameters $\lambda_1,\ldots,\lambda_M\in\mathbb{C}^+$, the standard $M$-lump solution exhibits normal scattering. Each of the $M$ lumps has a straight asymptotic trajectory with its own velocity and recovers its shape and velocity after the interaction, without a phase shift.

\subsection{A short review of the IST for the KP-I equation} \label{sec:22}

The Lax pair of the KP-I equation~\eqref{kp1} can be written as
\begin{equation} \label{lax-kp}
\begin{aligned}
    &(i\partial_y+\partial_{xx}+u)\psi=0, \\
    &(\partial_t+4\partial_{xxx}+6u\partial_x+3u_x-3i\partial_x^{-1}u_y)\psi=0.
\end{aligned}
\end{equation}
Under the transformation $\psi=\mu\exp\{i\kappa x-i\kappa^2y+4i\kappa^3t\}$, with spectral parameter $\kappa\in\mathbb{C}$, the spatial equation in the Lax pair~\eqref{lax-kp} becomes
\begin{equation} \label{lax2}
    (i\partial_y+\partial_{xx}+2i\kappa\partial_x+u)\mu=0,
\end{equation}
where the eigenfunction is normalized by the asymptotic condition $\mu=1+O(|\kappa|^{-1})$ as $|\kappa|\to\infty$. Expanding $\mu(\kappa)$ in a Laurent series about infinity,
\[
    \mu(\kappa)=1+\frac{\mu_1}{\kappa}+O(|\kappa|^{-2}), \quad |\kappa|\to\infty,
\]
and substituting this expansion into \eqref{lax2} yields the reconstruction formula
\begin{equation} \label{recover}
    u=-2i\partial_x\!\left[\lim_{|\kappa|\to\infty}\kappa(\mu-1)\right]=-2i\frac{\partial}{\partial x}\mu_{1}.
\end{equation}

Equation~\eqref{lax2} can be written in integral form \cite{KP-IST} as
\[
    \mu_\pm(x,y,\kappa)=1+G_\pm*(u\mu_\pm),
\]
where $*$ denotes convolution in the spatial variables, and the Green functions are given by
\[
G_\pm(x-x',y-y',\kappa)=\frac{i}{2\pi}\int_{\mathbb R}g_\pm(y-y',\zeta,\kappa)e^{i\zeta(x-x')}\,d\zeta,
\]
with
\begin{equation*}
    \begin{aligned}
        &g_+(y-y',\zeta,\kappa)=e^{-i\zeta(\zeta+2\kappa)(y-y')}\bigl[\Theta(y-y')\Theta(-\zeta)-\Theta(y'-y)\Theta(\zeta)\bigr],\\
        &g_-(y-y',\zeta,\kappa)=e^{-i\zeta(\zeta+2\kappa)(y-y')}\bigl[\Theta(y-y')\Theta(\zeta)-\Theta(y'-y)\Theta(-\zeta)\bigr].
    \end{aligned}
\end{equation*}
Here, $\Theta$ denotes the Heaviside function. The functions $g_+$ and $g_-$ are analytic in the upper and lower half-planes of $\kappa$, respectively. In general, the eigenfunctions $\mu_+$ and $\mu_-$ have different boundary values on the real $\kappa$-axis, with the jump determined by the scattering data. We consider reflectionless potentials for which these boundary values coincide and the eigenfunctions extend to a meromorphic function with finitely many poles. The corresponding potentials are pure lump solutions.

Assume that $\mu(\kappa)$ has $2M$ simple poles $\kappa_1,\ldots,\kappa_{2M}$. Near each pole, write
\begin{equation} \label{1pole}
    \mu(\kappa)=\frac{\phi}{\kappa-\kappa_j}+\nu_j(\kappa),
\end{equation}
where $\nu_j(\kappa)$ is the regular part at the pole $\kappa_j$. Substituting \eqref{1pole} into \eqref{lax2} and equating the coefficients of $(\kappa-\kappa_j)^{-1}$ and $(\kappa-\kappa_j)^0$, we obtain
\begin{equation*}
\begin{aligned}
&i\phi_y+\phi_{xx}+2i\kappa_j\phi_x+u\phi=0,\\
&i\nu_{j,y}+\nu_{j,xx}+2i\kappa_j\nu_{j,x}+u\nu_j+2i\phi_x=0.
\end{aligned}
\end{equation*}
In the second equation, $\nu_j$ and its derivatives are evaluated at $\kappa=\kappa_j$. Setting $\nu_j(\kappa_j)=m(x,y;\kappa_j)\phi$ and substituting into that equation gives
\[
    (im_y+m_{xx}+2i\kappa_jm_x)\phi+(2m_x+2i)\phi_x=0.
\]
For the standard lump reduction, choose $m_x=-i$ and $im_y+m_{xx}+2i\kappa_jm_x=0$. This gives
\[
    m=-i\left(x-2\kappa_jy+\tilde{\gamma}_j\right),
\]
where $\tilde\gamma_j$ is independent of $x$ and $y$. Its time dependence is determined by the temporal equation of the Lax pair.

For the standard $M$-lump solution, the eigenfunction is taken to be a sum over simple poles,
\[
    \mu(\kappa)=1+\sum_{\alpha=1}^{2M}\frac{\Phi_{\alpha}}{\kappa-\kappa_{\alpha}}.
\]
Evaluating the regular part at $\kappa_\alpha$,
\[
    \nu_\alpha(\kappa_\alpha)=m_\alpha\Phi_\alpha,\qquad \nu_\alpha(\kappa_\alpha)=\lim_{\kappa\to\kappa_\alpha}\left[\mu(\kappa)-\frac{\Phi_\alpha}{\kappa-\kappa_\alpha}\right],\quad \alpha=1,\ldots,2M,
\]
we obtain a closed linear system for the residues,
\[
    1+\sum_{\substack{\ell=1\\\ell\ne\alpha}}^{2M}\frac{\Phi_\ell}{\kappa_\alpha-\kappa_\ell}=m_\alpha\Phi_\alpha,\qquad \alpha=1,\ldots,2M,
\]
where $m_\alpha:=m(x,y;\kappa_\alpha)=-i\left(x-2\kappa_\alpha y+\tilde\gamma_\alpha\right)$. The temporal equation and the reconstruction formula~\eqref{recover} give
\[
    u=-2i\partial_x\sum_{\alpha=1}^{2M}\Phi_{\alpha}=2\frac{\partial^2}{\partial x^2}\ln{\tau},
\]
where $\tau=\tau_{2M}$ is the tau function appearing in \eqref{normal}, with $\lambda_j=-2\kappa_j$, $C_{j1}=\gamma_j$ and $\tilde\gamma_j=12\kappa_j^2t+\gamma_j$. A bounded, real-valued $M$-lump solution is obtained by imposing the conditions
\[
    \kappa_{j+M}=\kappa_j^*,\quad \gamma_{j+M}=\gamma_j^*, \quad j=1,2,\dots,M.
\]
The standard lump solution~\eqref{normal} obtained by the long-wave limit coincides with the IST solution under the identification $\lambda_j=-2\kappa_j$. We therefore also call $\lambda_j$ the spectral parameters of the lump solution.

Within the IST framework, multi-pole lump solutions arise when the meromorphic eigenfunction $\mu(\kappa)$ has higher-order poles. These poles can be obtained by coalescing distinct eigenvalues $\kappa_j$ while scaling the phase constants $\gamma_j$. This coalescence limit can be taken directly in the rational tau function in \eqref{normal}.

\subsection{Spectral coalescence of standard lump solutions}

The anomalous scattering of KP-I lumps was first discovered by Gorshkov, Pelinovsky, and Stepanyants \cite{JETP1993}, who applied the degenerate limit $\lambda_2\to\lambda_1$ to the standard two-lump solution~\eqref{normal} via a spectral perturbation.

For the standard two-lump solution, set
\[
    \lambda_1=\lambda,\quad \lambda_2=\lambda+\delta,\quad C_{11}=\frac{ic_1}{\delta},\quad C_{21}=\frac{ic_2}{\delta},
\]
where $\lambda\in\mathbb{C}^+$ and $\delta,c_1,c_2\in\mathbb{R}$. The associated tau function admits a Laurent expansion
\[
    \tau_4=\tau_{4,-4}\delta^{-4}+\tau_{4,-3}\delta^{-3}+\tau_{4,-2}\delta^{-2}+\tau_{4,-1}\delta^{-1}+\tau_{4,0}+O(\delta),\quad \delta\to0.
\]
To obtain a finite limit of the unnormalized tau function, we impose
\[
    \tau_{4,-4} = \tau_{4,-3} = \tau_{4,-2} = \tau_{4,-1} = 0.
\]
Solving these algebraic conditions yields two possible parameter relations:
\[
    \{c_1=2,\ c_2=-2\}\quad\text{or}\quad\{c_1=-2,\ c_2=2\}.
\]
Taking the limit $\delta\to 0$ then produces the degenerate two-lump solution
\[
    u=2\left(\ln\tau_{4,0}\right)_{xx}.
\]

More generally, applying spectral coalescence with suitable phase adjustments to the standard $M$-lump solution yields a degenerate $M$-lump solution comprising $M$ fundamental lumps. This procedure has two steps: first taking the long-wave limit of an $M$-lump chain, and then coalescing the spectral parameters of the resulting standard $M$-lump solution. In Section~\ref{sec:24}, we merge these two limiting processes into a single $\varepsilon\to 0$ limit.

\subsection{A generalized long-wave limit method with spectral perturbations} \label{sec:24}

In the classical construction, the spectral parameters $\lambda_j$ are pairwise distinct. The improved long-wave limit allows distinct proportionality factors in $k_j=\beta_j\varepsilon$ at a common spectral parameter \cite{anomalous-kp}, yielding the triangular three- and six-lump patterns. Here, we also introduce $O(\varepsilon)$ spectral perturbations and higher-order phase modulations.

Specifically, within the parameterization~\eqref{assume}, we prescribe the spectral parameters as
\begin{equation}\label{assume2}
 \lambda_j=\lambda+id_j\varepsilon,\qquad \lambda=a+bi,\quad b>0,\qquad j=1,\ldots,M,
\end{equation}
so that all $\lambda_j$ coalesce to a common value $\lambda\in\mathbb{C}^+$ as $\varepsilon\to 0$, with the real constants $d_j$ specifying the first-order spectral perturbations. We take $\beta_j$, $d_j$ and $C_{js}$ to be real, with $\beta_1=1$, $d_1=0$ and $C_{j1}=0$. We assume $\beta_jC_{j0}\ne0$ and exclude parameter values at which a nonremovable denominator in the interaction coefficients or the parameter constraints vanishes. The phase factors are written directly as
\[
 e^{\eta_{0,j}}=C_{j0}\exp\!\left(\sum_{s\geq2}C_{js}\varepsilon^s\right).
\]
Only finitely many $C_{js}$ are nonzero in the constructions below.

We expand $f_{2M}$ in powers of $\varepsilon$ up to order $M(M+1)$:
\[
 f_{2M}=\sum_{j=0}^{M(M+1)}f_{2M,j}\varepsilon^j+O\bigl(\varepsilon^{M(M+1)+1}\bigr).
\]
We then choose the parameters to annihilate all Taylor coefficients below the leading nonvanishing term of even order $2n$, writing
\[
 f_{2M}(\varepsilon)=\varepsilon^{2n}(\tau+\varepsilon R_\varepsilon),\qquad \tau=f_{2M,2n}>0.
\]
The coefficient $\tau$ determines the limiting solution. Here, an $M$th-order lump solution is one obtained from a $2M$-soliton tau function by this limiting procedure.

With the finite phase modulations used here, cancelling removable factors in the interaction coefficients gives a tau function analytic at $\varepsilon=0$. Taylor's theorem then bounds $R_\varepsilon$ and its required derivatives uniformly on compact sets of the physical variables. Hence, the normalized tau function
\[
 \widehat{f}_\varepsilon=\varepsilon^{-2n}f_{2M}(\varepsilon)=\tau+\varepsilon R_\varepsilon\longrightarrow\tau\quad\text{in }C^4_{\mathrm{loc}}(\mathbb R^3),
\]
with uniform convergence on compact sets of all $x,y,t$ derivatives of total order at most four. For a compact set $K$, let $m_K=\min_K\tau>0$. Then, $\widehat{f}_\varepsilon\geq m_K/2$ on $K$ for sufficiently small $|\varepsilon|$. Since the normalization is independent of $x,y,t$,
\[
u_\varepsilon=2(\ln f_{2M})_{xx}=2(\ln\widehat{f}_\varepsilon)_{xx}=2\frac{\widehat{f}_\varepsilon(\widehat{f}_\varepsilon)_{xx}-(\widehat{f}_\varepsilon)_x^2}{(\widehat{f}_\varepsilon)^2}\longrightarrow2\frac{\tau\tau_{xx}-\tau_x^2}{\tau^2}=2(\ln\tau)_{xx}.
\]
Write $\mathcal B[F]$ for the left-hand side of \eqref{kp-bilinear} with $f_N$ replaced by $F$. Quadratic homogeneity gives
\[
\mathcal B[\widehat{f}_\varepsilon]=\varepsilon^{-4n}\mathcal B[f_{2M}]=0.
\]
Passing to the limit in the bilinear equation gives $\mathcal B[\tau]=0$. Therefore, $u=2(\ln\tau)_{xx}$ solves the KP-I equation.

In the asymptotic propositions below, the lump indices $j$ label the individual peaks in each time limit ($t\to-\infty$ or $t\to+\infty$), all error bounds hold for fixed admissible parameters with $b>0$, and cube roots of real quantities denote their real branches.

\section{Second-order lump solutions} \label{sec:3}

For $M=2$, the asymptotic expansion of $f_4$ as $\varepsilon\to0$ takes the form
\[
    f_4=\sum_{j=0}^{6}f_{4,j}\,\varepsilon^j+O(\varepsilon^7).
\]
The second-order constructions below describe two or three asymptotically separated fundamental lumps. We treat these two cases separately and analyze their anomalous scattering.

\subsection{Anomalous scattering of second-order two-lump solutions} \label{sec:31}

For the second-order two-lump solution, we choose parameters such that
\[
    f_{4,0}=0,\quad f_{4,1}=0,\quad f_{4,2}=0,\quad f_{4,3}=0.
\]
The following two parameter branches satisfy these cancellation conditions:
\begin{equation*}
\begin{aligned}
    &\text{Case 1:}\quad\left\{C_{10}=\frac{1+\beta-d}{1-\beta+d},\quad C_{20}=\frac{1+\beta+d}{-1+\beta-d}\right\}, \\
    &\text{Case 2:}\quad\left\{C_{10}=\frac{1+\beta+d}{1-\beta-d},\quad C_{20}=\frac{1+\beta-d}{-1+\beta+d}\right\},
\end{aligned}
\end{equation*}
where $\beta:=\beta_2$ and $d:=d_2$ satisfy $\beta d\ne0$ (the degenerate case $d=0$ is deferred to Section~\ref{sec:32}). We use Case~1 below; the parameter formulas for Case~2 follow by replacing $d$ with $-d$. These parameters give $f_4=f_{4,4}\varepsilon^4+O(\varepsilon^5)$. Taking $\varepsilon\to0$ in Case~1 yields the second-order two-lump solution
\begin{equation} \label{21lump}
    u=2(\ln f_{4,4})_{xx}.
\end{equation}
The polynomial $f_{4,4}$ is given in Appendix~\ref{appendix-functions}.

\begin{figure}[ht]
\centering
\begin{minipage}{0.48\linewidth}\centering
\includegraphics[width=\linewidth]{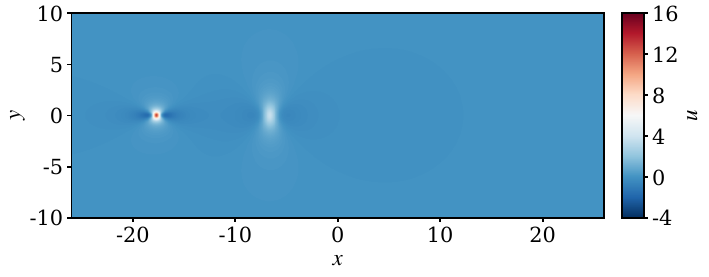}
\par\small (a1) $t=-2$.
\end{minipage}
\hfill
\begin{minipage}{0.48\linewidth}\centering
\includegraphics[width=\linewidth]{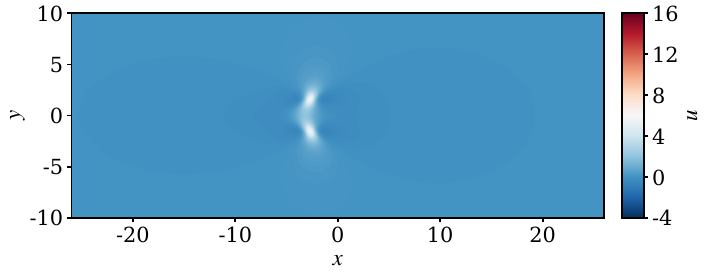}
\par\small (a2) $t=-0.5$.
\end{minipage}
\par\vspace{2mm}
\begin{minipage}{0.48\linewidth}\centering
\includegraphics[width=\linewidth]{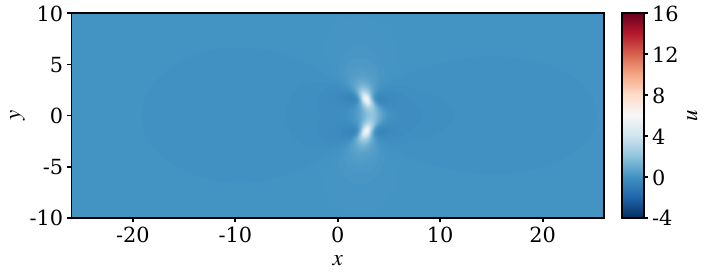}
\par\small (a3) $t=0.5$.
\end{minipage}
\hfill
\begin{minipage}{0.48\linewidth}\centering
\includegraphics[width=\linewidth]{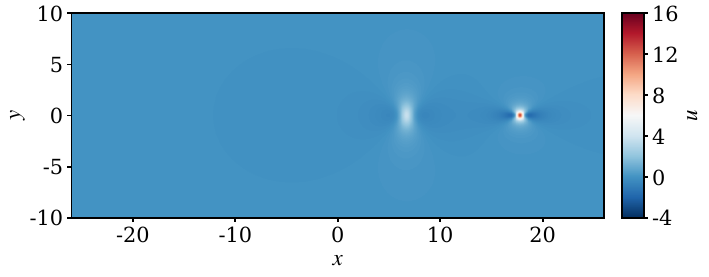}
\par\small (a4) $t=2$.
\end{minipage}
\caption{Normal scattering of the standard two-lump solution \eqref{normal} in the $(x,y)$-coordinates with $\lambda_1=i$, $\lambda_2=\frac74i$ and $C_{11}=C_{21}=0$.}
\label{normal-scatter}
\end{figure}

\begin{figure}[ht]
\centering
\begin{minipage}{0.48\linewidth}\centering
\includegraphics[width=\linewidth]{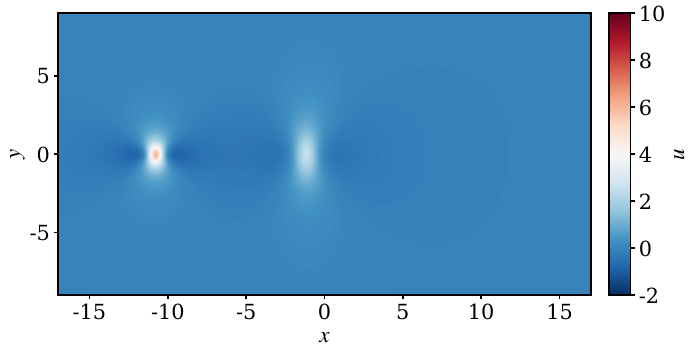}
\par\small (a1) $t=-2$.
\end{minipage}
\hfill
\begin{minipage}{0.48\linewidth}\centering
\includegraphics[width=\linewidth]{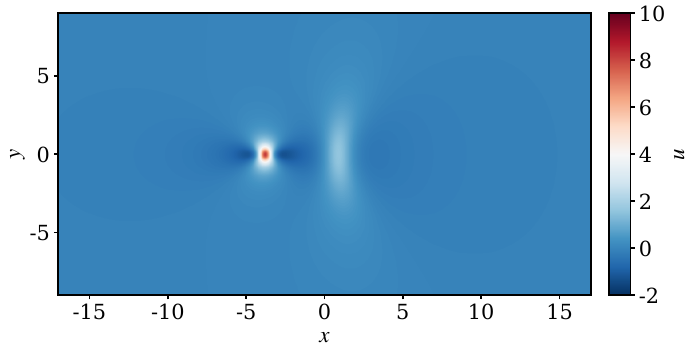}
\par\small (a2) $t=-0.5$.
\end{minipage}
\par\vspace{2mm}
\begin{minipage}{0.48\linewidth}\centering
\includegraphics[width=\linewidth]{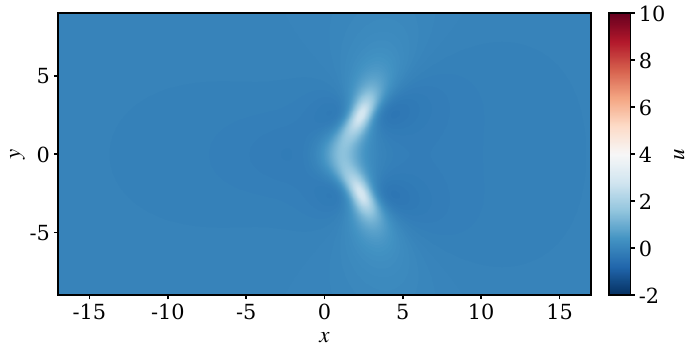}
\par\small (a3) $t=0.5$.
\end{minipage}
\hfill
\begin{minipage}{0.48\linewidth}\centering
\includegraphics[width=\linewidth]{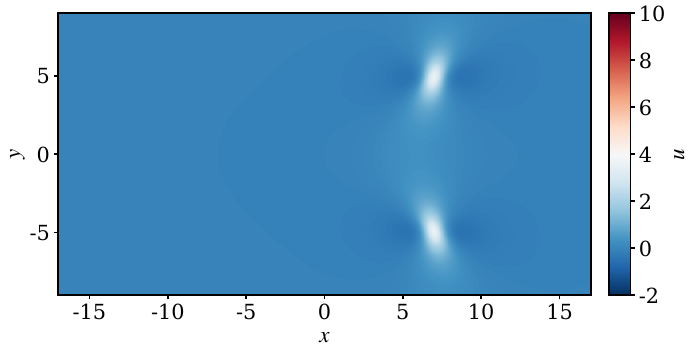}
\par\small (a4) $t=2$.
\end{minipage}
\caption{Anomalous scattering of the second-order two-lump solution \eqref{21lump} with $\lambda=i$, $\beta=d=1$ and $C_{12}=C_{22}=0$.}
\label{2lump-scatter}
\end{figure}

The standard lumps in Fig.~\ref{normal-scatter} have distinct asymptotic velocities and recover their shapes after the interaction. The degenerate pair in Fig.~\ref{2lump-scatter} shares a common asymptotic velocity and exhibits anomalous scattering. Proposition~\ref{anomalous21lump} gives its long-time positions and amplitudes.

\begin{proposition} \label{anomalous21lump}
The positions and amplitudes of the two lumps in \eqref{21lump} satisfy the following asymptotics.
\begin{enumerate}
    \item[(i)] As $t \to -\infty$,
    \begin{equation*}
        \begin{aligned}
        &x_1= 3(a^2+b^2)t-2\sqrt{3b|t|}+O(|t|^{-1/2}),\quad y_1= -6at+O(|t|^{-1/2}),\\
        &x_2= 3(a^2+b^2)t+2\sqrt{3b|t|}+O(|t|^{-1/2}),\quad y_2= -6at+O(|t|^{-1/2}),\\
        &A_1=4b^2+\frac{4\sqrt{b}}{\sqrt{3}}\sqrt{\frac{1}{|t|}}+O\bigl(|t|^{-1}\bigr),\quad A_2=4b^2-\frac{4\sqrt{b}}{\sqrt{3}}\sqrt{\frac{1}{|t|}}+O\bigl(|t|^{-1}\bigr),
        \end{aligned}
    \end{equation*}
    
    \item[(ii)] As $t \to +\infty$,
    \begin{equation*}
        \begin{aligned}
        &x_1= 3(a^2+b^2)t-\frac{2\sqrt{3}a}{\sqrt{b}}\sqrt{t}+\frac{1}{b}+O(t^{-1/2}),\quad y_1= -6at+\frac{2\sqrt{3}}{\sqrt{b}}\sqrt{t}+O(t^{-1/2}),\\
        &x_2= 3(a^2+b^2)t+\frac{2\sqrt{3}a}{\sqrt{b}}\sqrt{t}+\frac{1}{b}+O(t^{-1/2}),\quad y_2= -6at-\frac{2\sqrt{3}}{\sqrt{b}}\sqrt{t}+O(t^{-1/2}),\\
        &A_1=4b^2+O\bigl(t^{-1}\bigr),\quad A_2=4b^2+O\bigl(t^{-1}\bigr).
        \end{aligned}
    \end{equation*}
\end{enumerate}
\end{proposition}

\begin{proof}
Let $F$ be the polynomial in \eqref{appendix-f44} after removing its positive constant factor. In the coordinates $Z,Y$ of \eqref{appendix-coordinates}, set
\[
 q=\frac{2(C_{22}-\beta C_{12})}{\beta d},\qquad \tau_0=\xi^2+b^2\eta^2+b^{-2}.
\]
All parameters are fixed, with $b>0$ and the admissibility conditions of Section~\ref{sec:31}. As $t\to-\infty$, we set $\rho=|t|^{-1/2}$ and, for $s\in\{-1,1\}$, introduce the local coordinates $(\xi,\eta)$ via
\[
 Z=s\sqrt{12b}\,\rho^{-1}+\xi,\qquad Y=\eta.
\]
Direct expansion of \eqref{appendix-f44} gives
\[
 \frac{\rho^2F}{48b}=\tau_0+\rho P_{-,s}+O(\rho^2),\qquad P_{-,s}=\frac{s}{\sqrt{12b}}\left[\xi^3+b^2\xi\eta^2-2b\eta^2+(2b^{-2}-q)\xi+2b^{-3}\right].
\]
Similarly, as $t\to+\infty$, we set $\rho=t^{-1/2}$ and introduce
\[
 Z=b^{-1}+\xi,\qquad Y=s\sqrt{12/b}\,\rho^{-1}+\eta.
\]
Then
\[
\frac{\rho^2F}{48b}=\tau_0+\rho P_{+,s}+O(\rho^2),\qquad P_{+,s}=\frac{s\eta}{\sqrt{12/b}}(\tau_0-2\xi/b+q).
\]
Since the normalized tau functions are polynomials in $\rho$ and the local coordinates, the expansions and their derivatives of any fixed order hold uniformly on bounded sets of the local coordinates. The lower bound $\tau_0\geq b^{-2}$ allows expansion of the logarithm, giving
\[
 u=U_0+\rho U_{1,\pm,s}+O(\rho^2),\quad U_0=\frac{4(b^2\eta^2+b^{-2}-\xi^2)}{(\xi^2+b^2\eta^2+b^{-2})^2},\quad U_{1,\pm,s}=2\partial_\xi^2(P_{\pm,s}/\tau_0).
\]
At the origin,
\[
 U_0(0,0)=4b^2,\qquad \nabla U_0(0,0)=0,\qquad
 \nabla^2U_0(0,0)=\begin{pmatrix}-24b^4&0\\0&-8b^6\end{pmatrix}.
\]
The Hessian is negative definite. The implicit function theorem applied to the two stationary-point equations therefore gives a unique nearby critical point $(\xi,\eta)=O(\rho)$ for each branch, which remains a strict local maximum for sufficiently small $\rho$. Evaluating the field at this point gives
\[
 A=4b^2+\rho U_{1,\pm,s}(0,0)+O(\rho^2),\qquad
 U_{1,-,s}(0,0)=-4s\sqrt{b/3},\quad U_{1,+,s}(0,0)=0,
\]
because the displacement affects the amplitude only at order $\rho^2$.

At fixed time, the transformation to physical coordinates is invertible and preserves critical points and strict local maxima. Inverting it gives
\[
 x=3(a^2+b^2)t+Z-aY,\qquad y=-6at+Y.
\]
For negative time, $s=-1$ and $s=1$ give peaks~1 and~2, respectively; for positive time, $s=1$ and $s=-1$ give peaks~1 and~2. These substitutions yield the stated position remainders and amplitude corrections.
\qed
\end{proof}

The remaining asymptotic formulas follow from analogous local expansions and the same nondegenerate critical-point argument; the details are omitted.

\subsection{Anomalous scattering of second-order three-lump solutions} \label{sec:32}

The second-order three-lump solution is obtained by imposing
\[
    f_{4,0}=0,\quad f_{4,1}=0,\quad \cdots,\quad f_{4,5}=0.
\]
The following parameter constraints satisfy the cancellation conditions:
\[
    \left\{d=0, \quad C_{10}=-\frac{\beta+1}{\beta-1},\quad C_{20}=\frac{\beta+1}{\beta-1}, \quad C_{22}=\beta C_{12}\right\}.
\]
The choice $d=0$ gives $\lambda_1=\lambda_2=\lambda$. The coefficients $C_{12}$, $C_{13}$ and $C_{23}$ remain free in the phase expansion, but $C_{12}$ drops out of the limiting tau function. Under these constraints, $f_4=f_{4,6}\varepsilon^6+O(\varepsilon^7)$. Taking $\varepsilon\to0$ yields the second-order three-lump solution
\begin{equation}\label{22lump}
    u=2(\ln f_{4,6})_{xx}.
\end{equation}
The polynomial $f_{4,6}$ is given in Appendix~\ref{appendix-functions}.

\begin{figure}[ht]
\centering
\begin{minipage}{0.48\linewidth}\centering
\includegraphics[width=\linewidth]{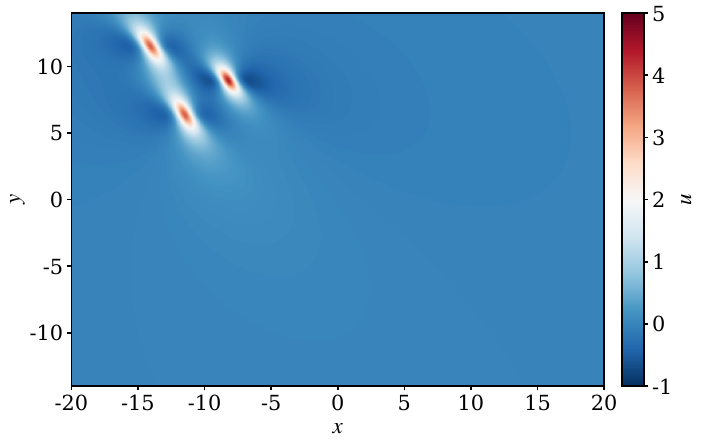}
\par\small (a1) $t=-3$.
\end{minipage}
\hfill
\begin{minipage}{0.48\linewidth}\centering
\includegraphics[width=\linewidth]{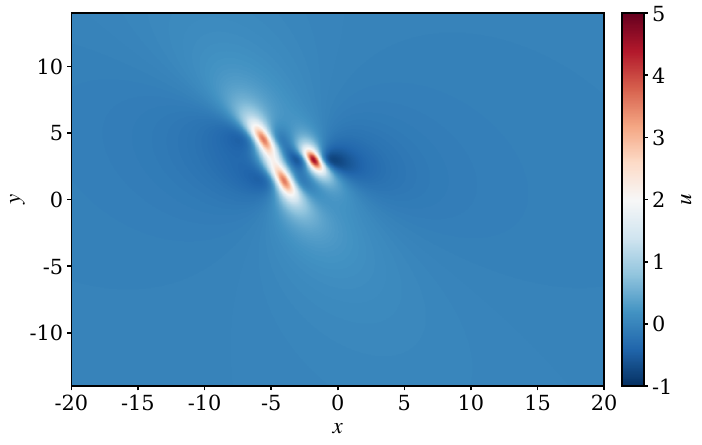}
\par\small (a2) $t=-1$.
\end{minipage}
\par\vspace{2mm}
\begin{minipage}{0.48\linewidth}\centering
\includegraphics[width=\linewidth]{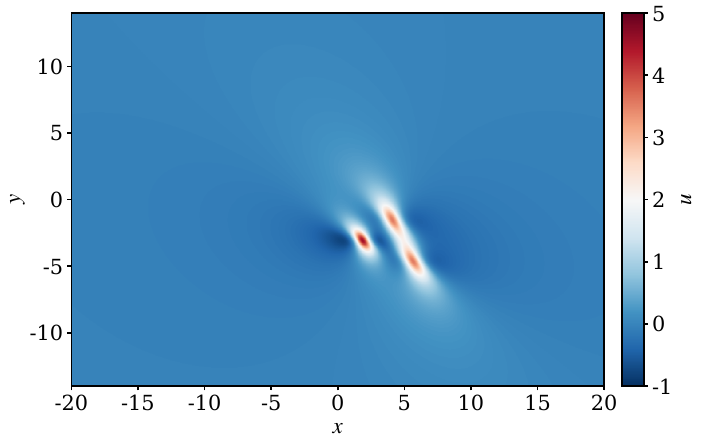}
\par\small (a3) $t=1$.
\end{minipage}
\hfill
\begin{minipage}{0.48\linewidth}\centering
\includegraphics[width=\linewidth]{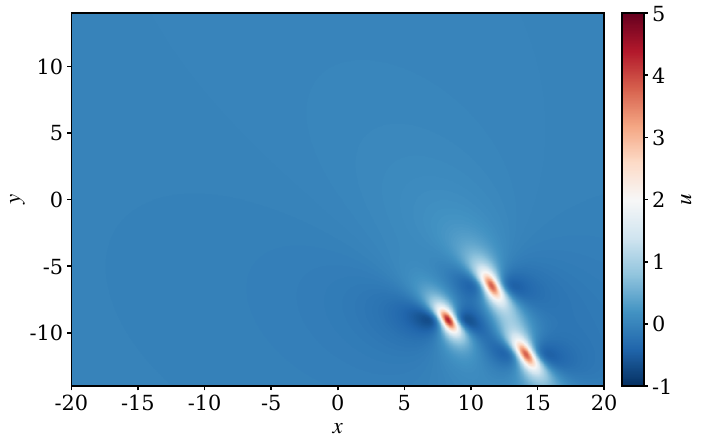}
\par\small (a4) $t=3$.
\end{minipage}
\caption{Evolution of the second-order three-lump solution \eqref{22lump} with $\lambda=\frac12+i$, $\beta=2$ and $C_{13}=C_{23}=0$.}
\label{22lump-scatter}
\end{figure}

Figure~\ref{22lump-scatter} shows the interaction of the second-order three-lump solution \eqref{22lump} for $C_{13}=C_{23}=0$. The separated lumps form a triangle in both time limits. In the co-moving tau function, $C_{13}$ and $C_{23}$ enter only through the time shift $t\mapsto t-(C_{23}-\beta C_{13})/[\beta(\beta^2-1)]$. Its long-time behavior is given below.

\begin{proposition}\label{anomalous3lump}
As $t\to\pm\infty$, the positions and amplitudes of the three lumps in \eqref{22lump} satisfy, for $j=1,2,3$,
\begin{align*}
&x_j=3(a^2+b^2)t+\left[\cos\frac{2(j-1)\pi}{3}-\frac ab\sin\frac{2(j-1)\pi}{3}\right]\sqrt[3]{-12t}+O(|t|^{-1/3}),\\
&y_j=-6at+\frac1b\sin\frac{2(j-1)\pi}{3}
 \sqrt[3]{-12t}+O(|t|^{-1/3}),\\
&A_j=4b^2+O(|t|^{-2/3}).
\end{align*}
\end{proposition}
Thus, $\beta$, $C_{13}$ and $C_{23}$ do not affect the displayed leading peak positions or the limiting amplitudes.

\subsection{Limits of lump chain solutions} \label{sec:33}

For $M=2$, we examine the degeneration of the lump chain as $\varepsilon$ decreases and the effects of $\beta$ and $d$ on the limit. Set
\[
    \eta_{0,1}=\ln{\frac{1+\beta-d}{1-\beta+d}}, \quad \eta_{0,2}=\ln{\frac{1+\beta+d}{-1+\beta-d}}.
\]
Then, the case $d\neq0$ corresponds to the situation analyzed in Section~\ref{sec:31}, whereas $d=0$ corresponds to that treated in Section~\ref{sec:32}.

\begin{figure}[ht]
\centering
\begin{minipage}{0.31\linewidth}\centering
\includegraphics[width=\linewidth]{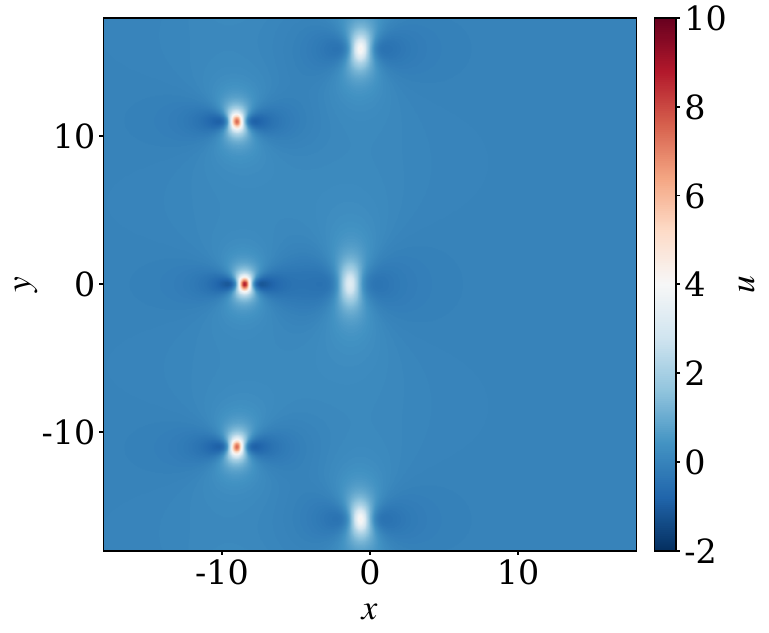}
\par\small (a1) $t=-1$.
\end{minipage}
\hfill
\begin{minipage}{0.31\linewidth}\centering
\includegraphics[width=\linewidth]{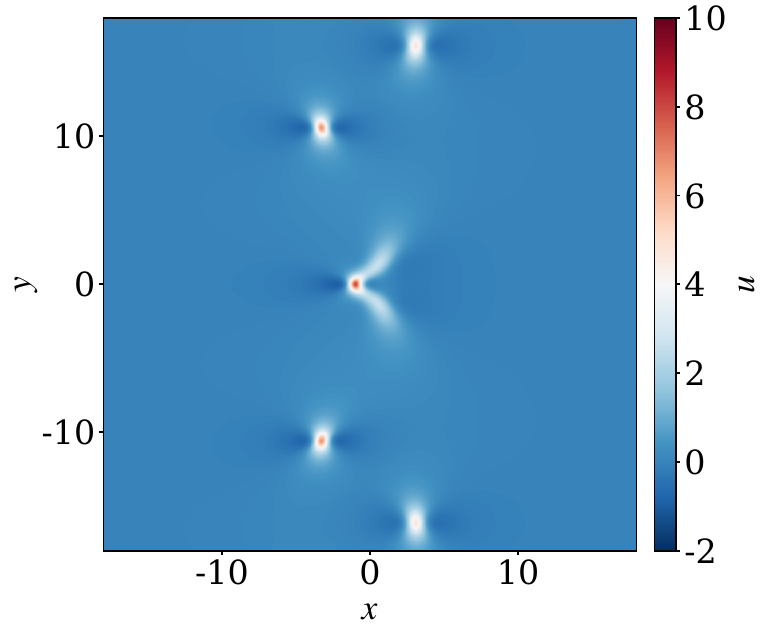}
\par\small (a2) $t=0$.
\end{minipage}
\hfill
\begin{minipage}{0.31\linewidth}\centering
\includegraphics[width=\linewidth]{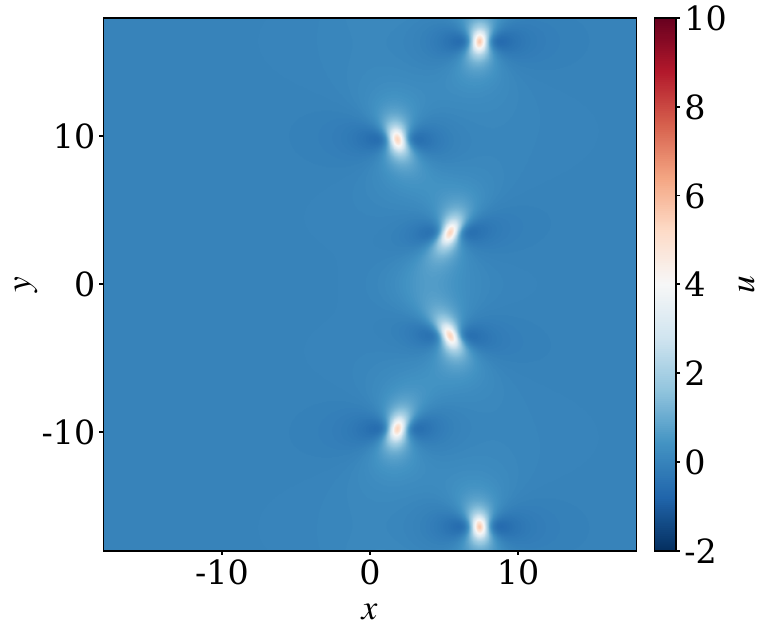}
\par\small (a3) $t=1$.
\end{minipage}
\par\vspace{2mm}
\begin{minipage}{0.31\linewidth}\centering
\includegraphics[width=\linewidth]{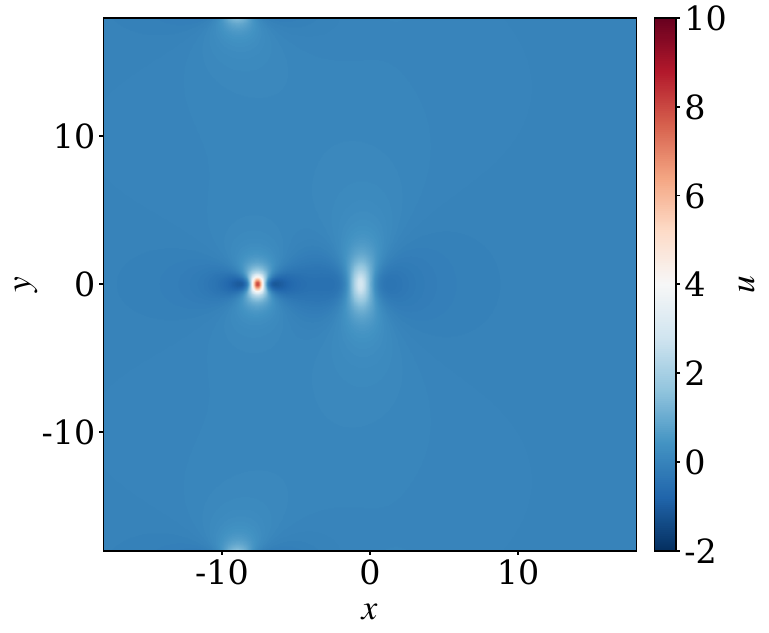}
\par\small (b1) $t=-1$.
\end{minipage}
\hfill
\begin{minipage}{0.31\linewidth}\centering
\includegraphics[width=\linewidth]{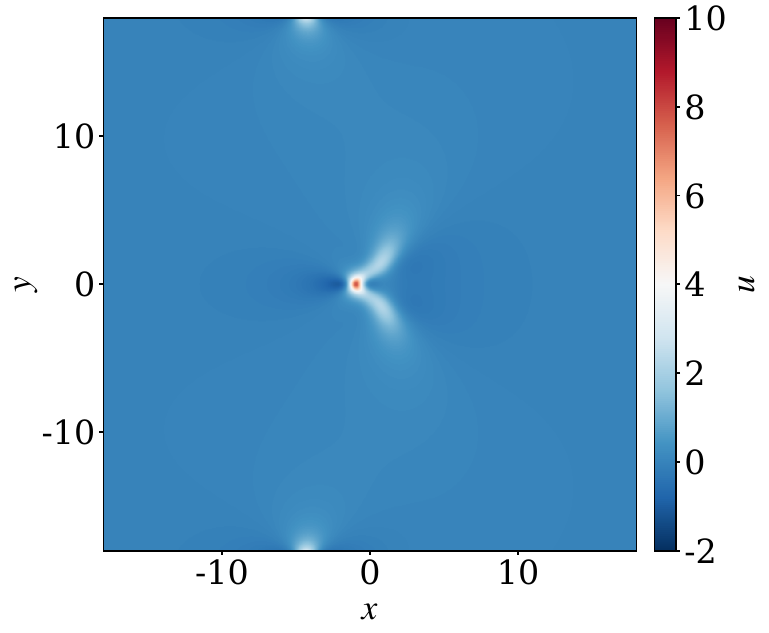}
\par\small (b2) $t=0$.
\end{minipage}
\hfill
\begin{minipage}{0.31\linewidth}\centering
\includegraphics[width=\linewidth]{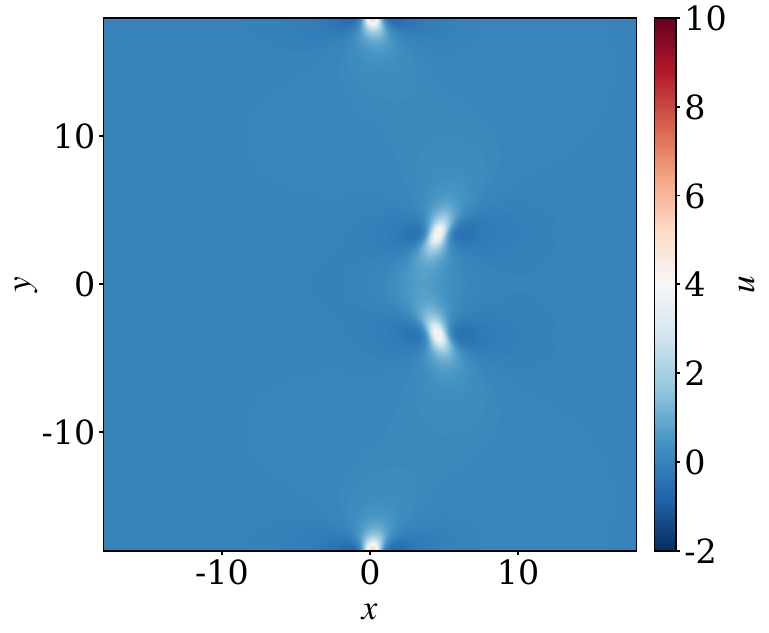}
\par\small (b3) $t=1$.
\end{minipage}
\caption{Evolution of the two-lump chain solution with $\lambda=i$, $\beta=1$ and $d=1$: upper row, $\varepsilon=\frac{2}{5}$; lower row, $\varepsilon=\frac{1}{4}$.}
\label{2LCd1}
\end{figure}

\begin{figure}[ht]
\centering
\begin{minipage}{0.31\linewidth}\centering
\includegraphics[width=\linewidth]{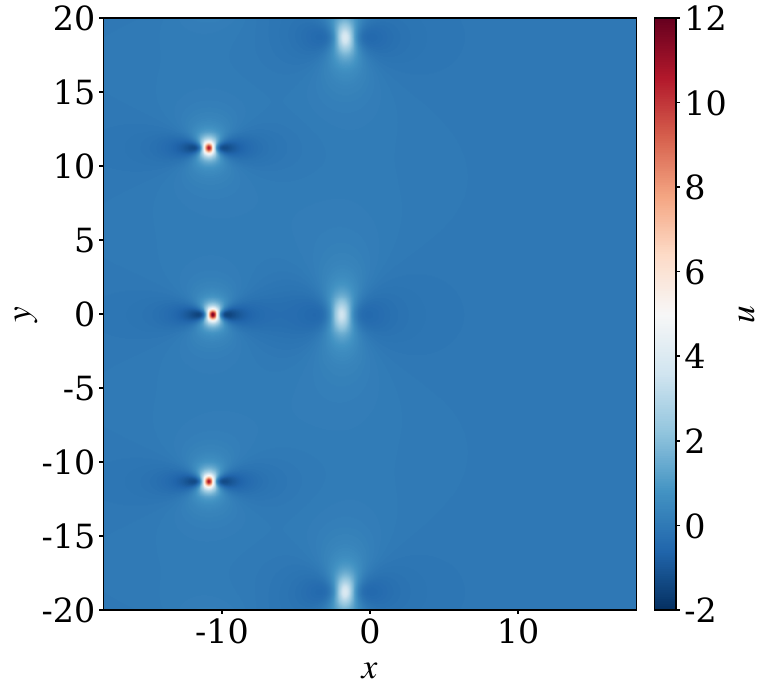}
\par\small (a1) $t=-1$.
\end{minipage}
\hfill
\begin{minipage}{0.31\linewidth}\centering
\includegraphics[width=\linewidth]{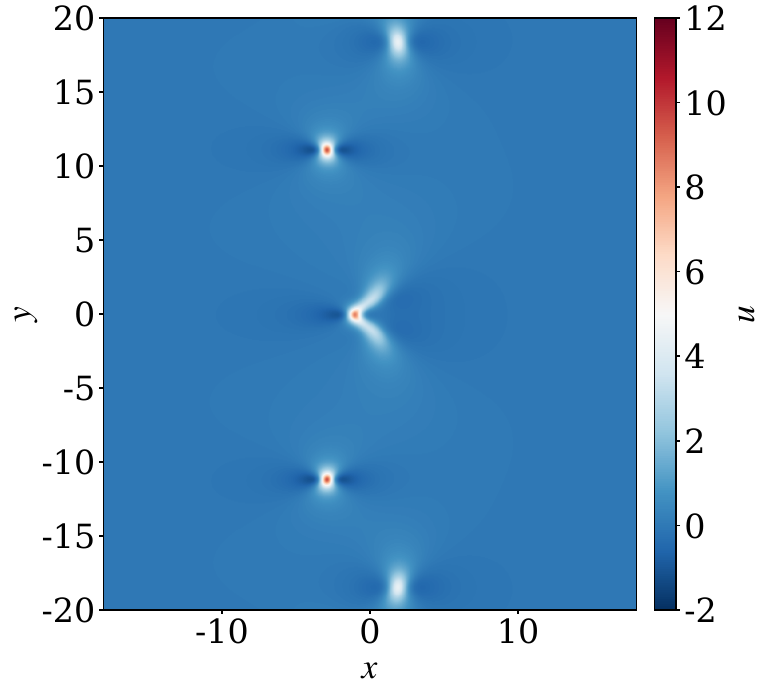}
\par\small (a2) $t=0$.
\end{minipage}
\hfill
\begin{minipage}{0.31\linewidth}\centering
\includegraphics[width=\linewidth]{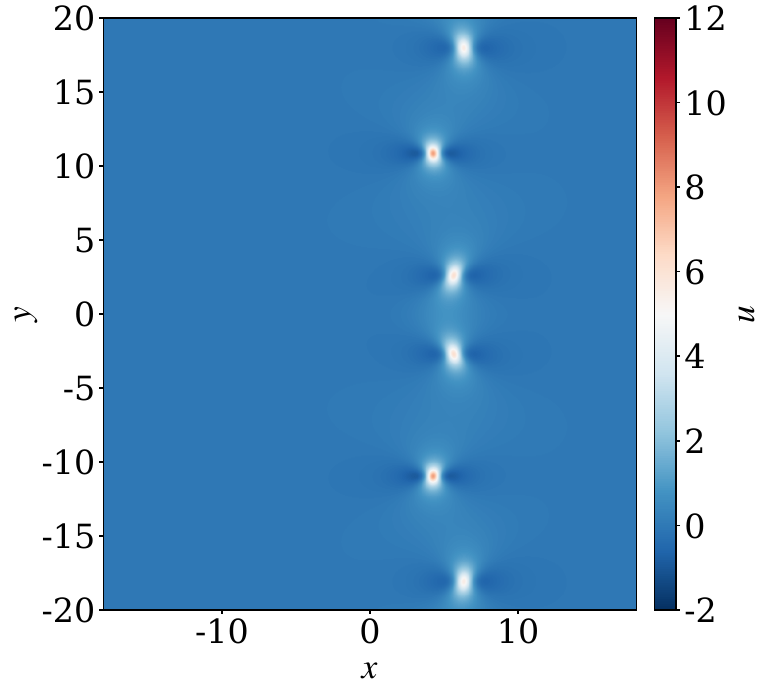}
\par\small (a3) $t=1$.
\end{minipage}
\caption{Evolution of the resonant two-lump chain solution with $\lambda=i$, $d=1$ and $\varepsilon=\frac13$, obtained as $\beta\to2$.}
\label{2LCd1-re}
\end{figure}

Figure~\ref{2LCd1} shows the degeneration of the two-lump chain into the second-order lump solution for $d\neq0$ and $\beta\neq1+d$, with $\lambda=i$. Two parallel lump chains undergo an elastic interaction around $t=0$. As $\varepsilon$ decreases, the spatial period increases. For fixed $t$, the solution in a bounded spatial window around the central lump unit near $y=0$ approaches \eqref{21lump}. For sufficiently small $\varepsilon$, only this central unit remains in the window.

When $d\neq0$, taking the resonant limit $\beta\to1+d$ produces a resonant two-lump chain solution. Multiplying $f_4$ by $(\beta-1-d)^2$ before taking the limit does not change $u$ and gives a finite, nonzero limiting tau function. As illustrated in Fig.~\ref{2LCd1-re}, the two chains undergo a resonant merger within the interaction region, while the parameter $\varepsilon$ continues to control the spatial period along the chain.

The $d=0$ degeneration of the two-lump chain, including its periodic triangular units, is discussed in Ref.~\cite{anomalous-kp}.

\begin{figure}[ht]
\centering
\begin{minipage}{0.31\linewidth}\centering
\includegraphics[width=\linewidth]{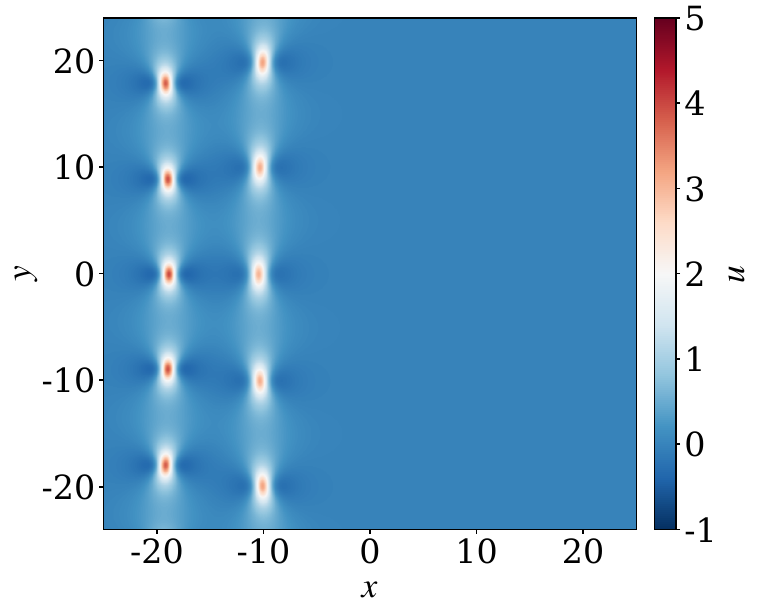}
\par\small (a1) $t=-4$.
\end{minipage}
\hfill
\begin{minipage}{0.31\linewidth}\centering
\includegraphics[width=\linewidth]{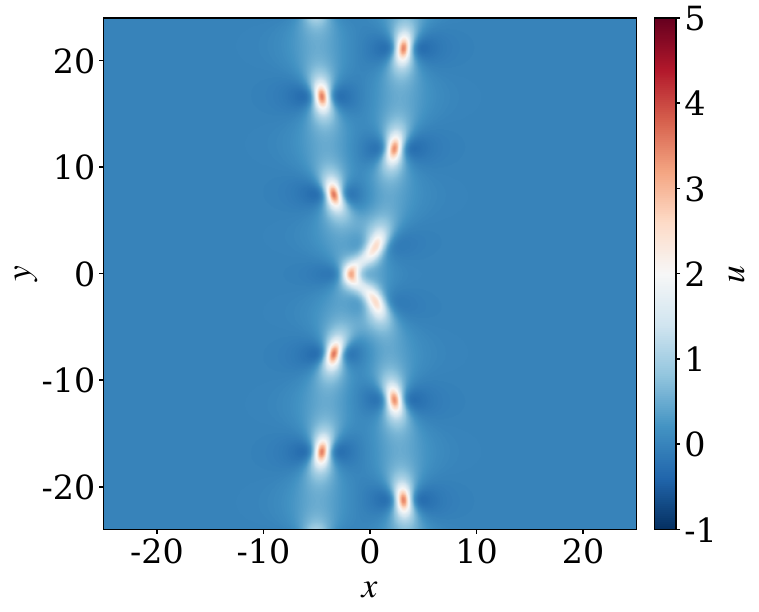}
\par\small (a2) $t=0$.
\end{minipage}
\hfill
\begin{minipage}{0.31\linewidth}\centering
\includegraphics[width=\linewidth]{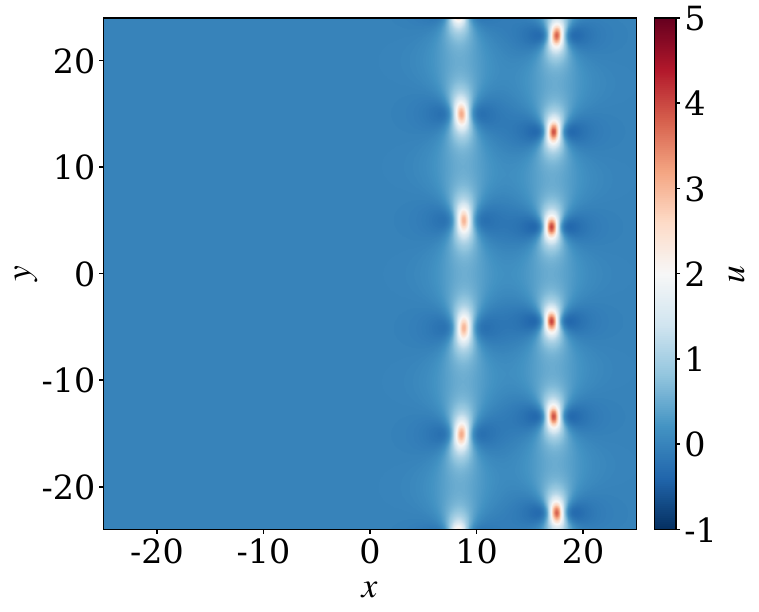}
\par\small (a3) $t=4$.
\end{minipage}
\caption{Evolution of the second-order lump chain solution with $\lambda=i$ and $\varepsilon=\frac{2}{3}$: the $\beta=1$, $d\to0$ case.}
\label{2ndLCd1}
\end{figure}

The two-lump chain can also degenerate into a second-order lump chain as $(\beta,d)$ approaches $(1,0)$. The joint limit $(\beta,d)\to(1,0)$ does not exist, so we consider the two iterated limits. 

First, fix $\beta=1$ and take $d\to0$ from the $d\neq0$ regime. This gives the second-order lump chain shown in Figure~\ref{2ndLCd1}. Near $t=0$, the two parallel chains interact weakly: the central lump unit undergoes a collision, while the other units follow smooth curved trajectories with little mutual interaction. The interaction between the chains is elastic. Taking $\varepsilon\to0$ then recovers the second-order lump solution \eqref{21lump}.

The other iterated limit, $d=0$ followed by $\beta\to1$, and the resulting second-order lump chain dynamics are discussed in Ref.~\cite{anomalous-kp}.

\section{Third-order lump solutions} \label{sec:4}

For $M=3$, the asymptotic expansion of $f_6$ as $\varepsilon\to0$ takes the form
\[
    f_6=\sum_{j=0}^{12}f_{6,j}\,\varepsilon^j+O(\varepsilon^{13}).
\]
The third-order solutions constructed below contain three, four, five or six asymptotically separated fundamental lumps. We give the corresponding parameter choices and analyze their anomalous scattering.

\subsection{Anomalous scattering of third-order three-lump solutions} \label{sec:41}

To construct the third-order three-lump solution, we require that the first six Taylor coefficients vanish:
\[
    f_{6,0}=0,\quad f_{6,1}=0,\quad \cdots,\quad f_{6,5}=0.
\]
Setting $\beta_2=\beta_3=1$, $d_2=1$, and $d_3=4$, we obtain four parameter branches, all of which satisfy $C_{12}=(4C_{22}-C_{32})/3$:
\begin{equation*}
\begin{aligned}
    &\text{Case 1a:}\quad\left\{C_{10}=\frac{1}{2},\quad C_{20}=-1,\quad C_{30}=-\frac{5}{2}\right\},\\
    &\text{Case 1b:}\quad\left\{C_{10}=-\frac{9}{2},\quad C_{20}=\frac{5}{3},\quad C_{30}=-\frac{1}{6}\right\}, \\
    &\text{Case 2a:}\quad\left\{C_{10}=-\frac{1}{2}-\sqrt{\frac{5}{2}},\quad C_{20}=\frac{1}{3}(-5+2\sqrt{10}),\quad C_{30}=\frac{1}{6}(-5-\sqrt{10})\right\}, \\
    &\text{Case 2b:}\quad\left\{C_{10}=-\frac{1}{2}+\sqrt{\frac{5}{2}},\quad C_{20}=\frac{1}{3}(-5-2\sqrt{10}),\quad C_{30}=\frac{1}{6}(-5+\sqrt{10})\right\}.
\end{aligned}
\end{equation*}
We use Case~1a below; Case~1b is analogous. Cases~2a and~2b reproduce the triangular three-lump family in Section~\ref{sec:32}. The first nonzero term of $f_6$ has order $\varepsilon^6$. Taking $\varepsilon\to0$ in Case~1a gives the third-order three-lump solution
\begin{equation}\label{31lump}
    u=2(\ln f_{6,6})_{xx}.
\end{equation}
The polynomial $f_{6,6}$ depends on $\lambda$ and on the phase parameters only through the combinations $C_{22}-C_{32}$ and $3C_{13}-4C_{23}+C_{33}$. Its expression is given in Appendix~\ref{appendix-functions}.

\begin{figure}[ht]
\centering
\begin{minipage}{0.48\linewidth}\centering
\includegraphics[width=\linewidth]{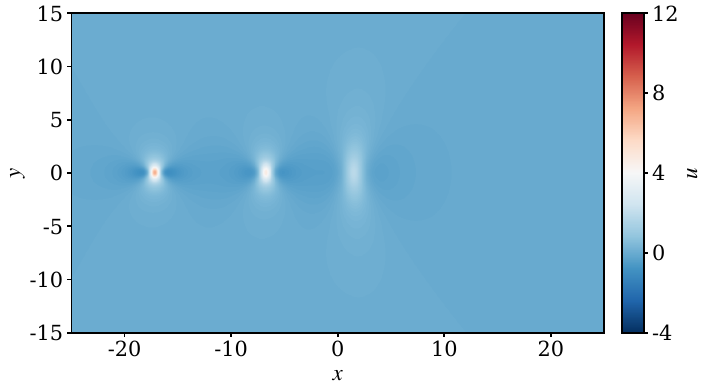}
\par\small (a1) $t=-2.5$.
\end{minipage}
\hfill
\begin{minipage}{0.48\linewidth}\centering
\includegraphics[width=\linewidth]{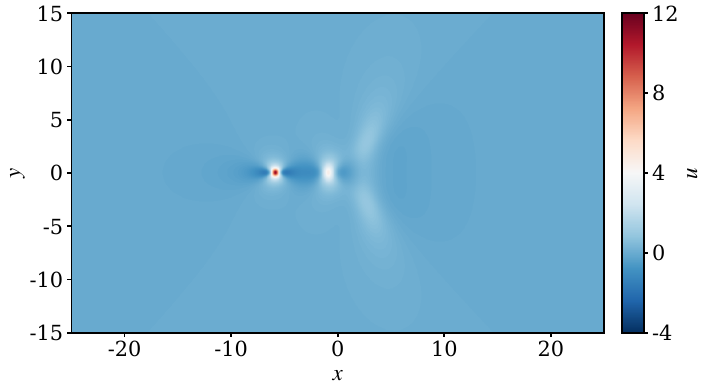}
\par\small (a2) $t=-0.5$.
\end{minipage}
\par\vspace{2mm}
\begin{minipage}{0.48\linewidth}\centering
\includegraphics[width=\linewidth]{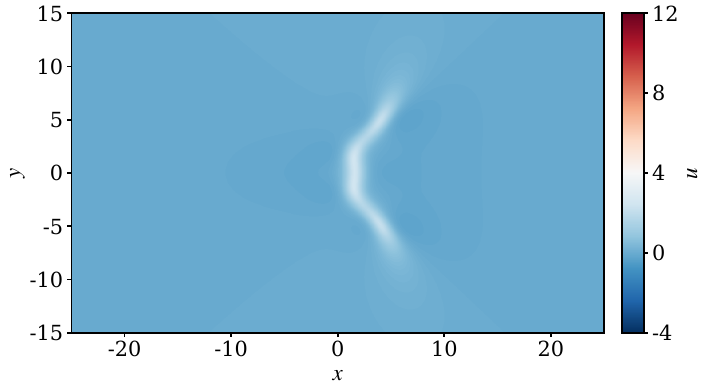}
\par\small (a3) $t=0.5$.
\end{minipage}
\hfill
\begin{minipage}{0.48\linewidth}\centering
\includegraphics[width=\linewidth]{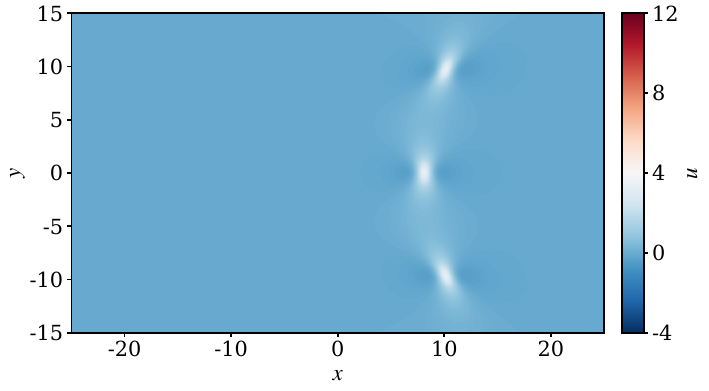}
\par\small (a4) $t=2.5$.
\end{minipage}
\caption{Evolution of the third-order three-lump solution \eqref{31lump} with $\lambda=i$ and $C_{22}=C_{32}=C_{13}=C_{23}=C_{33}=0$.}
\label{31lump-scatter}
\end{figure}

Figure~\ref{31lump-scatter} shows anomalous scattering of the three lumps. At large negative times, they share a common asymptotic velocity and their relative separations are of order $|t|^{1/2}$. The amplitudes vary during the interaction near $t=0$. The two outer outgoing lumps follow curved trajectories, while the central lump remains close to the common moving center. Proposition~\ref{anomalous31lump} gives the peak positions and amplitudes to the displayed orders; the phase parameters affect only the remainder terms.

\begin{proposition} \label{anomalous31lump}
The positions and amplitudes of the three lumps in \eqref{31lump} satisfy the following asymptotics.
\begin{enumerate}
    \item[(i)] As $t \to -\infty$,
    \begin{equation*}
        \begin{aligned}
        &x_1= 3(a^2+b^2)t-6\sqrt{b|t|}-\frac{1}{3b}+O(|t|^{-1/2}),\quad y_1= -6at+O(|t|^{-1/2}),\\
        &x_2= 3(a^2+b^2)t+\frac{2}{3b}+O(|t|^{-1}),\quad y_2= -6at+O(|t|^{-1}),\\
        &x_3= 3(a^2+b^2)t+6\sqrt{b|t|}-\frac{1}{3b}+O(|t|^{-1/2}),\quad y_3= -6at+O(|t|^{-1/2}),\\
        &A_1=4b^2+4\sqrt{\frac{b}{|t|}}+O\bigl(|t|^{-1}\bigr),\quad A_2=4b^2+O\bigl(|t|^{-1}\bigr),\quad A_3=4b^2-4\sqrt{\frac{b}{|t|}}+O\bigl(|t|^{-1}\bigr),
        \end{aligned}
    \end{equation*}
    
    \item[(ii)] As $t \to +\infty$,
    \begin{equation*}
        \begin{aligned}
        &x_1= 3(a^2+b^2)t-\frac{6a}{\sqrt{b}}\sqrt{t}+\frac{8}{3b}+O(t^{-1/2}),\quad y_1= -6at+\frac{6}{\sqrt{b}}\sqrt{t}+O(t^{-1/2}),\\
        &x_2= 3(a^2+b^2)t+\frac{2}{3b}+O(t^{-1}),\quad y_2= -6at+O(t^{-1}),\\
        &x_3= 3(a^2+b^2)t+\frac{6a}{\sqrt{b}}\sqrt{t}+\frac{8}{3b}+O(t^{-1/2}),\quad y_3= -6at-\frac{6}{\sqrt{b}}\sqrt{t}+O(t^{-1/2}),\\
        &A_1=4b^2+O\bigl(t^{-1}\bigr),\quad A_2=4b^2+O\bigl(t^{-1}\bigr),\quad A_3=4b^2+O\bigl(t^{-1}\bigr).
        \end{aligned}
    \end{equation*}
\end{enumerate}
\end{proposition}

\subsection{Anomalous scattering of third-order four-lump solutions}\label{sec:42}

To obtain the third-order four-lump solution, we require
\[
 f_{6,0}=0,\quad f_{6,1}=0,\quad\cdots,\quad f_{6,7}=0.
\]
Fixing $\beta_2=8$ and $\beta_3=11$, we find that these constraints are satisfied by the representative parameter set
\begin{equation}\label{four-parameters}
\begin{aligned}
 \bigg\{&\beta_1=1,\quad\beta_2=8,\quad\beta_3=11,\quad d_1=0,\quad d_2=1,\quad d_3=2,\\
 &C_{10}=-\frac53,\quad C_{20}=15,\quad C_{30}=-\frac{35}{2},\quad C_{js}=0\quad(j=1,2,3;\ s\geq1)\bigg\}.
\end{aligned}
\end{equation}
With the parameters in \eqref{four-parameters}, the expansion becomes
\[
 f_6=\varepsilon^8 f_{6,8}+O(\varepsilon^9).
\]
The long-wave limit gives the third-order four-lump solution
\begin{equation}\label{four-lump}
 u=2(\ln f_{6,8})_{xx},
\end{equation}
where $f_{6,8}$ is given in Appendix~\ref{appendix-functions}.

\begin{figure}[ht]
\centering
\begin{minipage}{0.48\linewidth}\centering
\includegraphics[width=\linewidth]{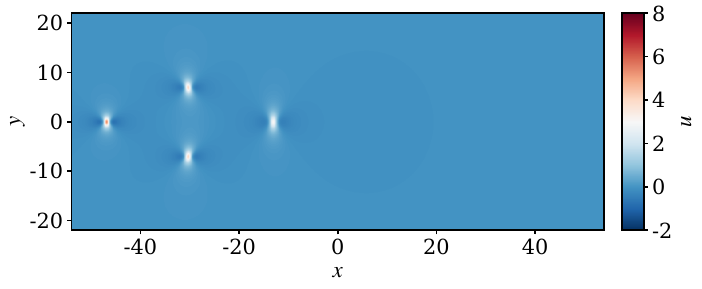}
\par\small (a1) $t=-10$.
\end{minipage}
\hfill
\begin{minipage}{0.48\linewidth}\centering
\includegraphics[width=\linewidth]{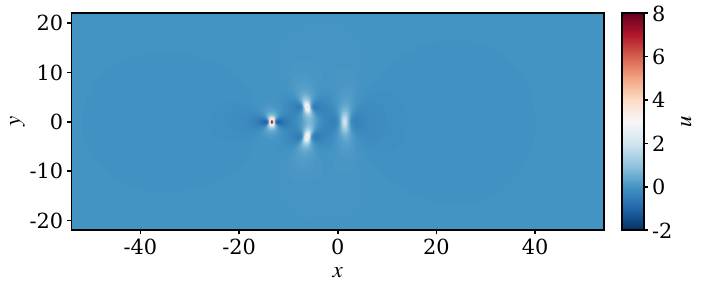}
\par\small (a2) $t=-2$.
\end{minipage}
\par\vspace{2mm}
\begin{minipage}{0.48\linewidth}\centering
\includegraphics[width=\linewidth]{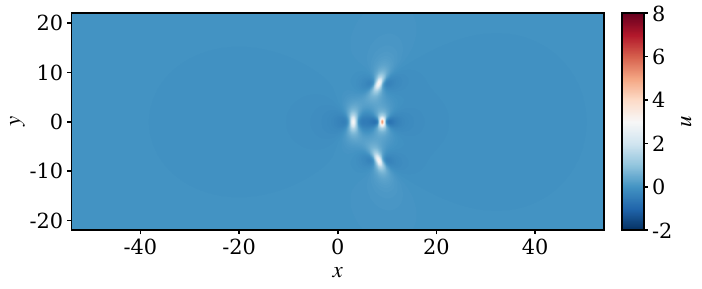}
\par\small (a3) $t=2$.
\end{minipage}
\hfill
\begin{minipage}{0.48\linewidth}\centering
\includegraphics[width=\linewidth]{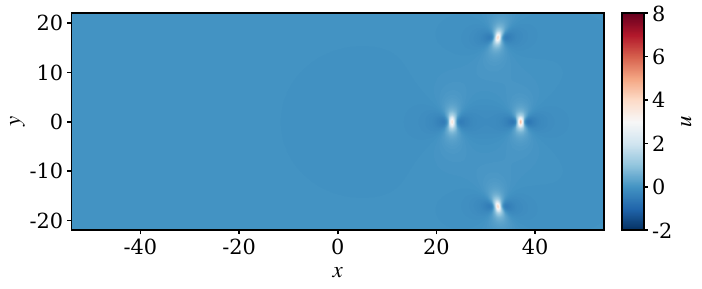}
\par\small (a4) $t=10$.
\end{minipage}
 \caption{Evolution of the third-order four-lump solution~\eqref{four-lump} with $\lambda=i$ and the parameters~\eqref{four-parameters}.}
 \label{four-scatter}
\end{figure}

Its long-time behavior is summarized below.

\begin{proposition}\label{four-asymptotics}
The positions and amplitudes of the four lumps in~\eqref{four-lump} satisfy the following asymptotics.
\begin{enumerate}
 \item[(i)] As $t\to-\infty$,
 \begin{equation*}
 \begin{aligned}
 &x_1=3(a^2+b^2)t+\sqrt{12b(\sqrt2+1)|t|}+O(|t|^{-1/2}),\quad y_1=-6at+O(|t|^{-1/2}),\\
 &x_2=3(a^2+b^2)t-\sqrt{12b(\sqrt2+1)|t|}+O(|t|^{-1/2}),\quad y_2=-6at+O(|t|^{-1/2}),\\
 &x_3=3(a^2+b^2)t-a\sqrt{\frac{12(\sqrt2-1)|t|}{b}}+\frac{1-\sqrt2}{b}+O(|t|^{-1/2}),\quad y_3=-6at+\sqrt{\frac{12(\sqrt2-1)|t|}{b}}+O(|t|^{-1/2}),\\
 &x_4=3(a^2+b^2)t+a\sqrt{\frac{12(\sqrt2-1)|t|}{b}}+\frac{1-\sqrt2}{b}+O(|t|^{-1/2}),\quad y_4=-6at-\sqrt{\frac{12(\sqrt2-1)|t|}{b}}+O(|t|^{-1/2}), \\
 &A_1=4b^2-\frac23\sqrt{\frac{12b(\sqrt2+1)}{|t|}}+O(|t|^{-1}),\quad A_2=4b^2+\frac23\sqrt{\frac{12b(\sqrt2+1)}{|t|}}+O(|t|^{-1}), \\
 &A_3=4b^2+O(|t|^{-1}),\quad A_4=4b^2+O(|t|^{-1}).
 \end{aligned}
 \end{equation*}
 
 \item[(ii)] As $t\to+\infty$,
 \begin{equation*}
 \begin{aligned}
 &x_1=3(a^2+b^2)t+\sqrt{12b(\sqrt2-1)t}+O(t^{-1/2}),\quad y_1=-6at+O(t^{-1/2}),\\
 &x_2=3(a^2+b^2)t-\sqrt{12b(\sqrt2-1)t}+O(t^{-1/2}),\quad y_2=-6at+O(t^{-1/2}),\\
 &x_3=3(a^2+b^2)t-a\sqrt{\frac{12(\sqrt2+1)t}{b}}+\frac{1+\sqrt2}{b}+O(t^{-1/2}),\quad y_3=-6at+\sqrt{\frac{12(\sqrt2+1)t}{b}}+O(t^{-1/2}),\\
 &x_4=3(a^2+b^2)t+a\sqrt{\frac{12(\sqrt2+1)t}{b}}+\frac{1+\sqrt2}{b}+O(t^{-1/2}),\quad y_4=-6at-\sqrt{\frac{12(\sqrt2+1)t}{b}}+O(t^{-1/2}), \\
 &A_1=4b^2+\frac23\sqrt{\frac{12b(\sqrt2-1)}{t}}+O(t^{-1}),\quad A_2=4b^2-\frac23\sqrt{\frac{12b(\sqrt2-1)}{t}}+O(t^{-1}),\\
 &A_3=4b^2+O(t^{-1}),\quad A_4=4b^2+O(t^{-1}).
 \end{aligned}
 \end{equation*}
\end{enumerate}
\end{proposition}

Figure~\ref{four-scatter} shows the anomalous scattering of the four lumps. For $\lambda=i$, the incoming configuration has a longer horizontal pair and a shorter vertical pair relative to the common center $(3t,0)$. After the interaction, their separation lengths interchange, and the leading configuration is rotated by $\pi/2$. For general $\lambda$, the trajectories in Proposition~\ref{four-asymptotics} give the corresponding deformation in the original coordinates. All four amplitudes approach $4b^2$, and the four lumps share the asymptotic velocity $(3(a^2+b^2),-6a)$, with relative separations of order $|t|^{1/2}$.

\subsection{Anomalous scattering of third-order five-lump solutions}\label{sec:43}

To obtain the third-order five-lump solution, we set the first ten Taylor coefficients to zero:
\[
 f_{6,0}=0,\quad f_{6,1}=0,\quad\cdots,\quad f_{6,9}=0.
\]
With $\beta_2=1$ and $\beta_3=7$, these conditions are fulfilled by the parameter choice
\begin{equation}\label{five-parameters}
\begin{aligned}
 \bigg\{&\beta_1=1,\quad\beta_2=1,\quad\beta_3=7,\quad d_1=0,\quad d_2=8,\quad d_3=4,\\
 &C_{10}=-\frac32,\quad C_{20}=-\frac32,\quad C_{30}=-\frac{12}{5},\quad C_{js}=0\quad(j=1,2,3;\ s\geq1)\bigg\}.
\end{aligned}
\end{equation}
With the parameters in \eqref{five-parameters}, we obtain
\[
 f_6=\varepsilon^{10}f_{6,10}+O(\varepsilon^{11}).
\]
The corresponding third-order five-lump solution is
\begin{equation}\label{five-lump}
 u=2(\ln f_{6,10})_{xx},
\end{equation}
with $f_{6,10}$ given in Appendix~\ref{appendix-functions}.

\begin{figure}[ht]
 \centering
 \begin{minipage}{0.48\linewidth}\centering
\includegraphics[width=\linewidth]{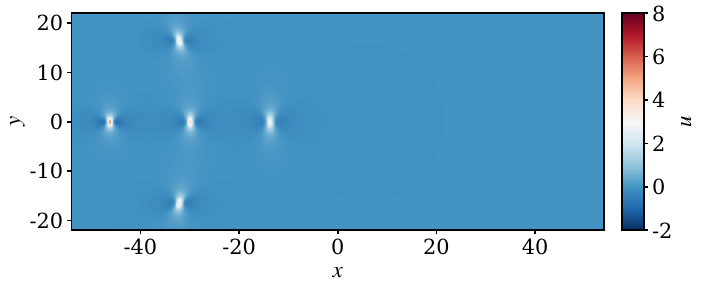}
\par\small (a1) $t=-10$.
\end{minipage}
\hfill
\begin{minipage}{0.48\linewidth}\centering
\includegraphics[width=\linewidth]{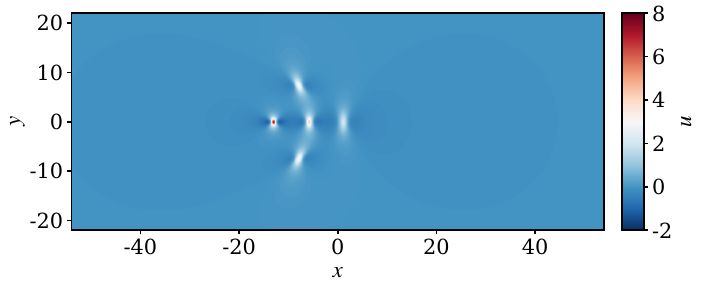}
\par\small (a2) $t=-2$.
\end{minipage}
\par\vspace{2mm}
\begin{minipage}{0.48\linewidth}\centering
\includegraphics[width=\linewidth]{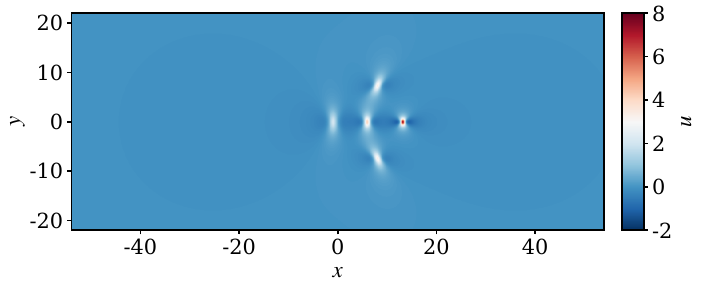}
\par\small (a3) $t=2$.
\end{minipage}
\hfill
\begin{minipage}{0.48\linewidth}\centering
\includegraphics[width=\linewidth]{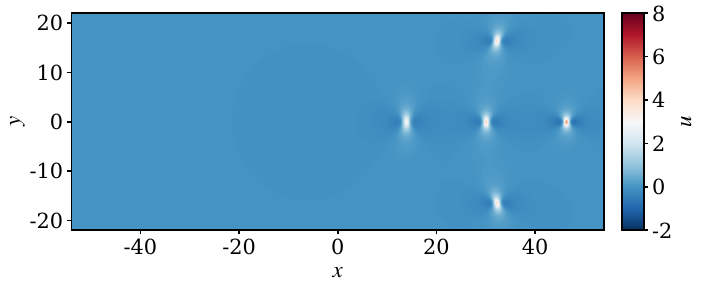}
\par\small (a4) $t=10$.
\end{minipage}
 \caption{Evolution of the third-order five-lump solution~\eqref{five-lump} with $\lambda=i$ and the parameters~\eqref{five-parameters}.}
 \label{five-scatter}
\end{figure}

\Needspace{16\baselineskip}
\begin{proposition}\label{five-asymptotics}
The positions and amplitudes of the five lumps in~\eqref{five-lump} satisfy the following asymptotics.
\begin{enumerate}
 \item[(i)] As $t\to-\infty$,
 \begin{equation*}
 \begin{aligned}
 &x_1=3(a^2+b^2)t+\sqrt{12\sqrt5\,b|t|}+O(|t|^{-1/2}),\quad y_1=-6at+O(|t|^{-1/2}),\\
 &x_2=3(a^2+b^2)t-\sqrt{12\sqrt5\,b|t|}+O(|t|^{-1/2}),\quad y_2=-6at+O(|t|^{-1/2}),\\
 &x_3=3(a^2+b^2)t-a\sqrt{\frac{12\sqrt5\,|t|}{b}}-\frac{\sqrt5}{b}+O(|t|^{-1/2}),\quad y_3=-6at+\sqrt{\frac{12\sqrt5\,|t|}{b}}+O(|t|^{-1/2}),\\
 &x_4=3(a^2+b^2)t+a\sqrt{\frac{12\sqrt5\,|t|}{b}}-\frac{\sqrt5}{b}+O(|t|^{-1/2}), \quad y_4=-6at-\sqrt{\frac{12\sqrt5\,|t|}{b}}+O(|t|^{-1/2}),\\
 &x_5=3(a^2+b^2)t-\frac{1}{3b^4t}+O(|t|^{-2}),\quad y_5=-6at+O(|t|^{-2}), \\
 &A_1=4b^2-\frac23\sqrt{\frac{12\sqrt5\,b}{|t|}}+O(|t|^{-1}),\quad A_2=4b^2+\frac23\sqrt{\frac{12\sqrt5\,b}{|t|}}+O(|t|^{-1}),\\
 &A_3=4b^2+O(|t|^{-1}),\quad A_4=4b^2+O(|t|^{-1}),\quad A_5=4b^2+O(|t|^{-2}).
 \end{aligned}
 \end{equation*}
 
 \item[(ii)] As $t\to+\infty$,
 \begin{equation*}
 \begin{aligned}
 &x_1=3(a^2+b^2)t+\sqrt{12\sqrt5\,bt}+O(t^{-1/2}),\quad y_1=-6at+O(t^{-1/2}),\\
 &x_2=3(a^2+b^2)t-\sqrt{12\sqrt5\,bt}+O(t^{-1/2}),\quad y_2=-6at+O(t^{-1/2}),\\
 &x_3=3(a^2+b^2)t-a\sqrt{\frac{12\sqrt5\,t}{b}}+\frac{\sqrt5}{b}+O(t^{-1/2}),\quad y_3=-6at+\sqrt{\frac{12\sqrt5\,t}{b}}+O(t^{-1/2}),\\
 &x_4=3(a^2+b^2)t+a\sqrt{\frac{12\sqrt5\,t}{b}}+\frac{\sqrt5}{b}+O(t^{-1/2}),\quad y_4=-6at-\sqrt{\frac{12\sqrt5\,t}{b}}+O(t^{-1/2}),\\
 &x_5=3(a^2+b^2)t-\frac{1}{3b^4t}+O(t^{-2}),\quad y_5=-6at+O(t^{-2}), \\
 &A_1=4b^2+\frac23\sqrt{\frac{12\sqrt5\,b}{t}}+O(t^{-1}),\quad A_2=4b^2-\frac23\sqrt{\frac{12\sqrt5\,b}{t}}+O(t^{-1}),\\
 &A_3=4b^2+O(t^{-1}),\quad A_4=4b^2+O(t^{-1}),\quad A_5=4b^2+O(t^{-2}).
 \end{aligned}
 \end{equation*}
\end{enumerate}
\end{proposition}

Figure~\ref{five-scatter} shows four outer lumps surrounding a central lump. For $\lambda=i$, the four outer positions have the same leading square configuration as $t\to-\infty$ and $t\to+\infty$, while the horizontal positions of the vertical pair change from $3t-\sqrt5+O(|t|^{-1/2})$ to $3t+\sqrt5+O(t^{-1/2})$. The opposite amplitude corrections on the horizontal pair change sign between the two time limits. The central lump approaches the common moving center. Thus, the long-time configuration retains its leading square pattern, while the trajectories and amplitude corrections describe the anomalous scattering. All five limiting amplitudes are $4b^2$.

\subsection{Anomalous scattering of third-order six-lump solutions} \label{sec:44}

Constructing the third-order six-lump solution requires the first twelve Taylor coefficients to vanish:
\[
    f_{6,0}=0,\quad f_{6,1}=0,\quad \cdots,\quad f_{6,11}=0.
\]
For $\beta_2=2$ and $\beta_3=3$, the cancellation conditions are satisfied by
\begin{equation*}
\begin{aligned}
    \bigg\{&d_2=0, \quad d_3=0, \quad C_{10}=-6,\quad C_{20}=15, \quad C_{30}=-10, \\ 
    &C_{12}=\frac{1}{3}C_{32}, \quad C_{22}=\frac{2}{3}C_{32}, \quad C_{13}=\frac{1}{5}(4C_{23}-C_{33}),\quad C_{14}=\frac{1}{5}(4C_{24}-C_{34})\bigg\}.
\end{aligned}
\end{equation*}
These parameters give
\[
 f_6=\varepsilon^{12}f_{6,12}+O(\varepsilon^{13}).
\]
The long-wave limit yields the third-order six-lump solution
\begin{equation}\label{33lump}
 u=2(\ln f_{6,12})_{xx},
\end{equation}
whose auxiliary function is listed in Appendix~\ref{appendix-functions}. In the co-moving coordinates, the phase parameters $C_{23}$, $C_{33}$, $C_{15}$, $C_{25}$ and $C_{35}$ enter $f_{6,12}$ only through the effective time $T$ and the parameter $H$:
\[
 T=t+\frac{3C_{23}-2C_{33}}{30},\qquad H=5C_{15}-4C_{25}+C_{35}.
\]
The common translation in physical coordinates remains $(3(a^2+b^2)t,-6at)$. For fixed phase parameters, $T-t$ is constant and does not change the leading long-time asymptotics.

\begin{figure}[ht]
\centering
\begin{minipage}{0.48\linewidth}\centering
\includegraphics[width=\linewidth]{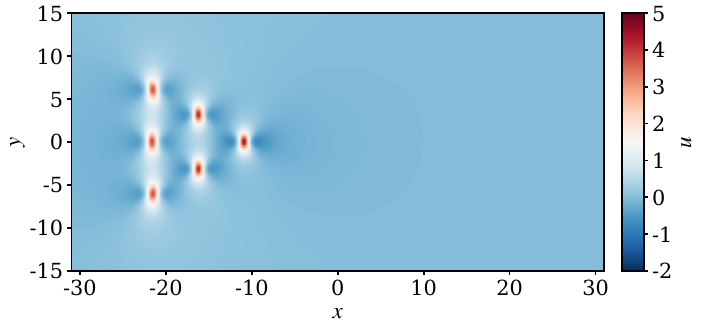}
\par\small (a1) $t=-6$.
\end{minipage}
\hfill
\begin{minipage}{0.48\linewidth}\centering
\includegraphics[width=\linewidth]{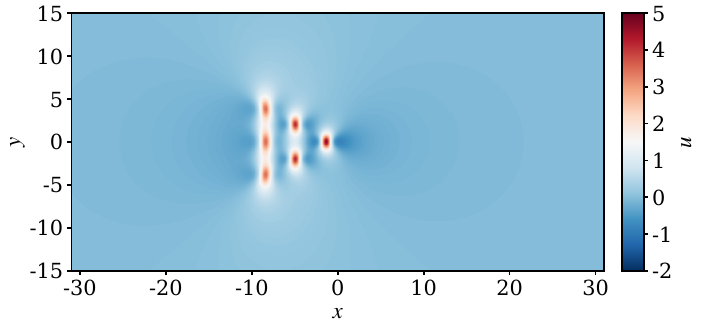}
\par\small (a2) $t=-2$.
\end{minipage}
\par\vspace{2mm}
\begin{minipage}{0.48\linewidth}\centering
\includegraphics[width=\linewidth]{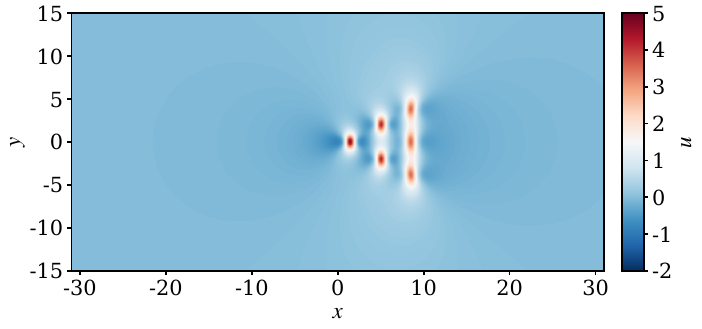}
\par\small (a3) $t=2$.
\end{minipage}
\hfill
\begin{minipage}{0.48\linewidth}\centering
\includegraphics[width=\linewidth]{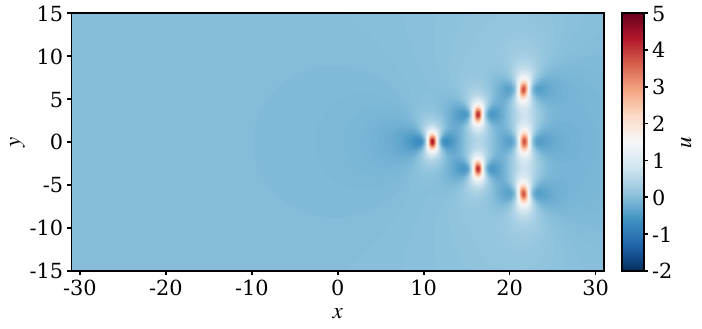}
\par\small (a4) $t=6$.
\end{minipage}
\par\vspace{2mm}
\begin{minipage}{0.48\linewidth}\centering
\includegraphics[width=\linewidth]{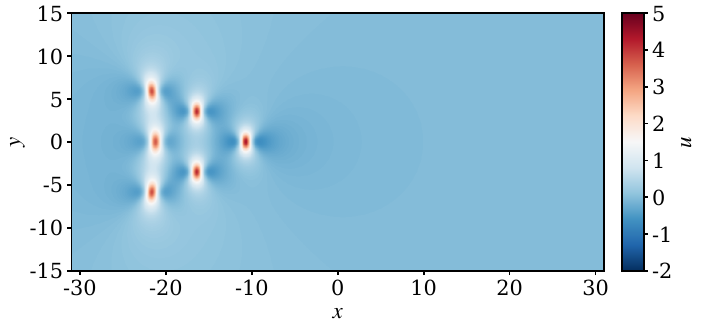}
\par\small (b1) $t=-6$.
\end{minipage}
\hfill
\begin{minipage}{0.48\linewidth}\centering
\includegraphics[width=\linewidth]{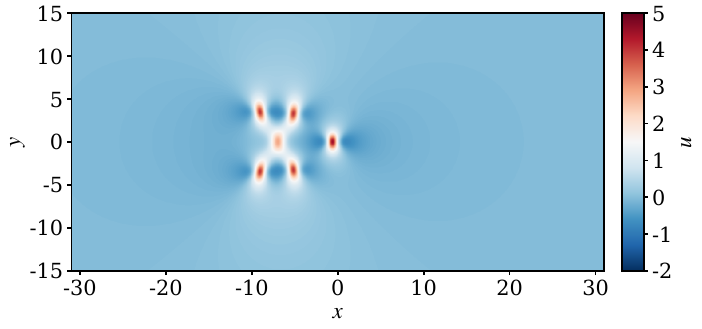}
\par\small (b2) $t=-2$.
\end{minipage}
\par\vspace{2mm}
\begin{minipage}{0.48\linewidth}\centering
\includegraphics[width=\linewidth]{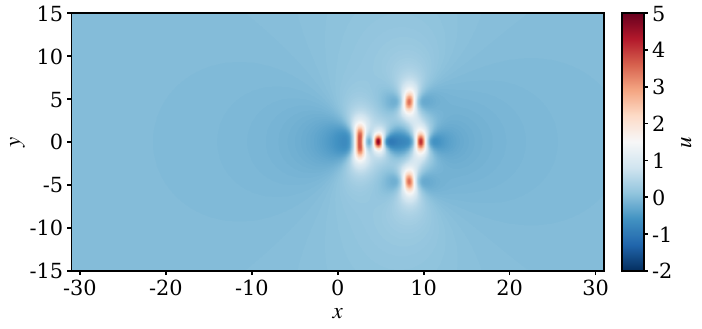}
\par\small (b3) $t=2$.
\end{minipage}
\hfill
\begin{minipage}{0.48\linewidth}\centering
\includegraphics[width=\linewidth]{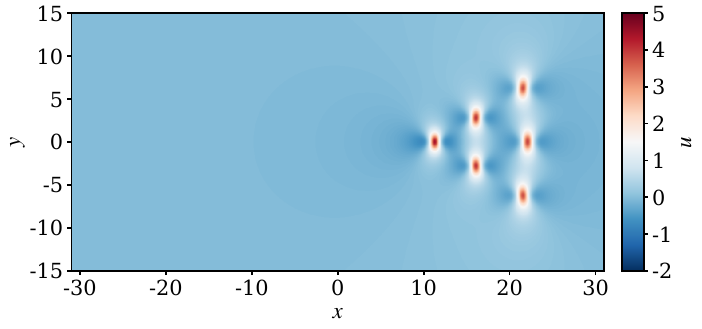}
\par\small (b4) $t=6$.
\end{minipage}
\caption{Evolution of the third-order six-lump solution \eqref{33lump} with $\lambda=i$ and $C_{23}=C_{33}=0$: (a1)--(a4), $C_{15}=C_{25}=C_{35}=0$; (b1)--(b4), $C_{25}=100$, $C_{15}=C_{35}=0$.}
\label{33lump-scatter}
\end{figure}

Figure~\ref{33lump-scatter} shows the evolution for $T=t$ with $H=0$ and $H=-400$. For large $|t|$, the six fundamental lumps form two triangles, in agreement with the long-time patterns in Ref.~\cite{yjkkp}. The effective phase combinations affect the finite-time interaction while leaving this leading asymptotic configuration unchanged.

\begin{proposition}\label{anomalous33lump-time}
For fixed phase parameters, the peak positions of the six-lump solution \eqref{33lump} satisfy, as $t\to\pm\infty$ and for $j=1,2,3$,
\begin{align*}
&x_j=3(a^2+b^2)t+\left[\cos\frac{2j\pi}{3}-\frac ab\sin\frac{2j\pi}{3}\right]\sqrt[3]{-(30+18\sqrt5)t}+O(|t|^{-1/3}),\\
&y_j=-6at+\frac1b\sin\frac{2j\pi}{3}\sqrt[3]{-(30+18\sqrt5)t}+O(|t|^{-1/3}),\\
&x_{j+3}=3(a^2+b^2)t+\left[\cos\frac{2j\pi}{3}-\frac ab\sin\frac{2j\pi}{3}\right]\sqrt[3]{(18\sqrt5-30)t}+O(|t|^{-1/3}),\\
&y_{j+3}=-6at+\frac1b\sin\frac{2j\pi}{3}
 \sqrt[3]{(18\sqrt5-30)t}+O(|t|^{-1/3}).
\end{align*}
Their amplitudes satisfy $A_j=4b^2+O(|t|^{-2/3})$, $j=1,\ldots,6$.
\end{proposition}

At $T=0$, the internal parameter $H$ produces the following large-parameter configuration.

\begin{proposition}\label{anomalous33lump}
Assume $\lambda=i$ and $T=0$, and let $\theta_j=(2j-1)\pi/5$, $j=1,\ldots,5$. As $H\to+\infty$, the peak positions of the six-lump solution \eqref{33lump} satisfy
\begin{align*}
 &x_j=3t+(6H)^{1/5}\cos\theta_j+O(H^{-1/5}),\\
 &y_j=(6H)^{1/5}\sin\theta_j+O(H^{-1/5}),\qquad j=1,\ldots,5,\\
 &x_6=3t-\frac{10}{H}+O(H^{-2}),\qquad y_6=0,
\end{align*}
and amplitudes
\[
 A_j=4+O(H^{-2/5}),\qquad j=1,\ldots,5,\qquad A_6=4+O(H^{-2}).
\]
\end{proposition}

\begin{figure}[ht]
\centering
\begin{minipage}{0.48\linewidth}\centering
\includegraphics[width=\linewidth]{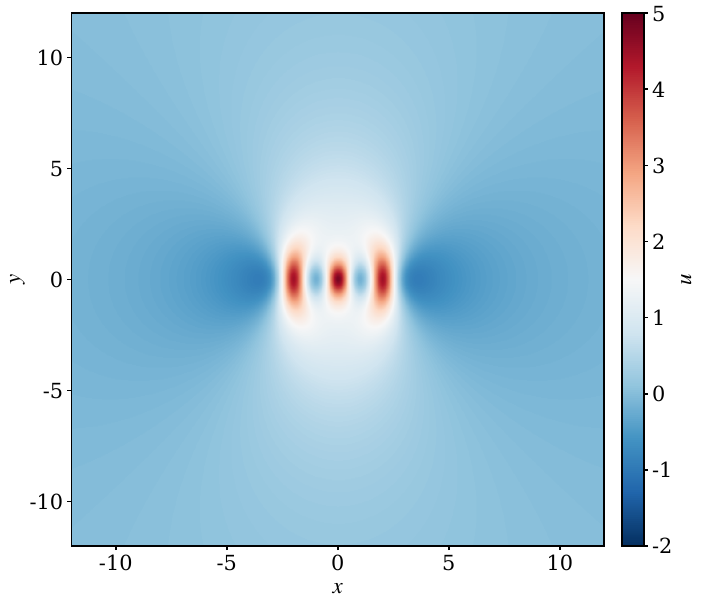}
\par\small (a)
\end{minipage}
\hfill
\begin{minipage}{0.48\linewidth}\centering
\includegraphics[width=\linewidth]{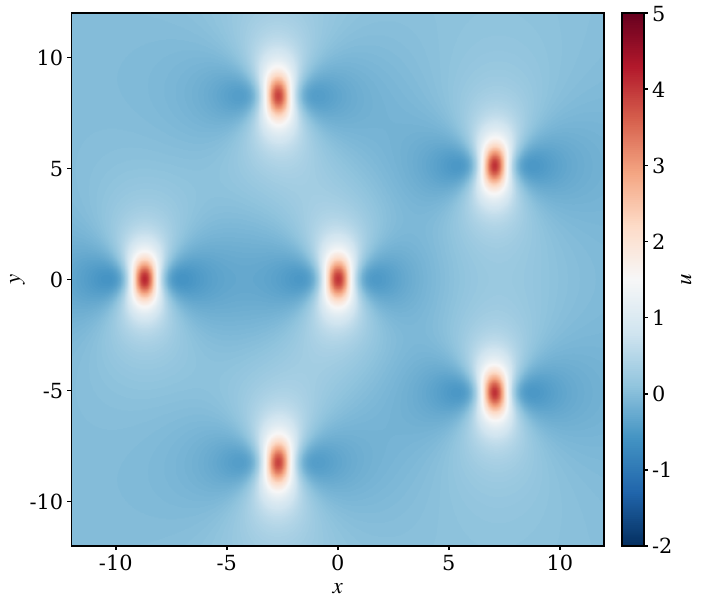}
\par\small (b)
\end{minipage}
\caption{Configurations of the third-order six-lump solution \eqref{33lump} with $\lambda=i$ and $T=0$: (a) $H=0$; (b) $H=10000$. The standard representative $C_{23}=C_{33}=C_{15}=C_{25}=0$, $C_{35}=H$, $t=0$ is used in both panels.}
\label{33lump-large-para}
\end{figure}

For the representative used in Fig.~\ref{33lump-large-para}, $T=0$ implies $t=0$, so the common translation vanishes. Other phase choices with the same $T$ and $H$ give the same co-moving solution, with the physical translation determined by $t$. We call the zero-phase configuration at $T=H=0$ the ground state. At $T=0$ and $\lambda=i$, the solution separates, as $H\to+\infty$, into one central lump and five outer lumps whose leading positions form a regular pentagon.

\section{Discussion and conclusions} \label{sec:5}

We have developed a generalized long-wave limit method for the multi-soliton tau function by introducing wavenumber rescalings, spectral perturbations, and higher-order phase modulations through a single small parameter $\varepsilon$. For KP-I, the construction yields second-order two- and three-lump solutions and third-order three-, four-, five-, and six-lump solutions. It gives a unified derivation of these solutions, including the triangular three- and six-lump solutions in \cite{anomalous-kp}.

We have derived the long-time asymptotics of the peak positions and amplitudes, together with the sublinear separation rates for the solutions in Sections~\ref{sec:3} and~\ref{sec:4}, and described the degeneration of lump chains into localized higher-order lumps. For the third-order six-lump solution, the phase parameters enter through an effective time $T$ and an internal parameter $H$. At $\lambda=i$ and $T=0$, the limit $H\to+\infty$ gives one central lump and five outer lumps whose leading positions form a regular pentagon.

The degenerate lump solutions constructed here correspond to members of the families classified by Chakravarty and Zowada \cite{class-KP} and are related to the multi-pole lump solutions obtained by Villarroel and Ablowitz through the IST \cite{multipole}. The present method obtains these solutions directly from the multi-soliton tau function.

An extension of the present method is to divide the spectral parameters into $L$ clusters with pairwise distinct limiting values $\lambda^{(1)},\ldots,\lambda^{(L)}\in\mathbb{C}^+$, replacing \eqref{assume2} by
\begin{equation*}
\begin{aligned}
    &\lambda_j=\lambda^{(1)}+id_j^{(1)}\varepsilon,\quad j=1,\ldots,M_1,\\
&\lambda_j=\lambda^{(2)}+id_j^{(2)}\varepsilon,\quad j=M_1+1,\ldots,M_1+M_2,\\
&\qquad\vdots\\
&\lambda_j=\lambda^{(L)}+id_j^{(L)}\varepsilon,\quad j=1+\sum_{l=1}^{L-1}M_l,\ldots,\sum_{l=1}^{L}M_l,
\end{aligned}
\end{equation*}
where $M_1+\cdots+M_L=M$ and $M_l$ is the construction order of the $l$th cluster. This parameterization suggests a construction of hybrid higher-order lump solutions, with normal scattering between clusters and anomalous scattering within each cluster. Further extensions include complex spectral and phase perturbations, applications to other $(2+1)$-dimensional integrable equations, and analogous limits of breather solutions for constructing higher-order rogue waves.

\par\medskip

\noindent \textbf{Acknowledgements} This work is supported by the Kaiyuan International Mathematical Sciences Institute, the Fundamental Research Funds for the Central Universities (JKF2025077492962), and the Academic Excellence Foundation of BUAA for Ph.D. Students.

During the preparation of this manuscript, the authors used OpenAI GPT-6.1 Sol and GPT-6 Astra to assist with mathematical derivations. They reviewed, verified, and edited all AI-generated content and take full responsibility for its correctness and for the final manuscript.

\section*{Declarations}
\noindent \textbf{Conflict of interest} The authors declare that they have no conflict of interest.

\noindent\textbf{Data availability} The computational materials supporting this work are not publicly available. Requests for further information should be addressed to the authors.

\appendix
\section{Complete auxiliary functions}\label{appendix-functions}
\setcounter{equation}{0}
\renewcommand{\theequation}{A.\arabic{equation}}
We use the co-moving coordinates
\begin{equation}\label{appendix-coordinates}
X=x-3(a^2+b^2)t,\qquad Y=y+6at,\qquad Z=X+aY.
\end{equation}
The auxiliary function for the lump solution in \eqref{21lump} is
\begin{equation}\label{appendix-f44}
f_{4,4}=\frac{\beta^2d^2}{(1-\beta+d)^2}\Biggl\{\left[Z^2-b^2Y^2+12bt-\frac{2(C_{22}-\beta C_{12})}{\beta d}\right]^2+4Y^2(bZ-1)^2+\frac4{b^2}\left(Z+\frac1b\right)^2+4Y^2+\frac4{b^4}\Biggr\}.
\end{equation}
The auxiliary function for the lump solution in \eqref{22lump} is
\begin{equation*}
\begin{aligned}
f_{4,6}=\frac{\beta^2(\beta+1)^2}{36}\Biggl\{&\left[Z^3-3b^2ZY^2+12t-\frac{12(C_{23}-\beta C_{13})}{\beta(\beta^2-1)}\right]^2\\
 &+b^2Y^2(3Z^2-b^2Y^2)^2+\frac9{b^2}(Z^2+b^2Y^2)^2+\frac{36}{b^2}Y^2+\frac{36}{b^6}\Biggr\}.
\end{aligned}
\end{equation*}
The auxiliary function for the lump solution in \eqref{31lump} is
\begin{equation*}
\begin{aligned}
f_{6,6}={}&\left[Z^3-3b^2ZY^2+6bY^2+36btZ-24t+2(C_{22}-C_{32})Z+2C_{13}-\frac83C_{23}+\frac23C_{33}\right]^2\\
 &+b^2Y^2\left[3Z^2-b^2Y^2-\frac{6Z}{b}+36bt+2(C_{22}-C_{32})\right]^2\\
 &+\frac9{b^2}\left\{\left[Z^2-b^2Y^2+12bt+\frac{2Z}{b}+\frac2{b^2}+\frac23(C_{22}-C_{32})\right]^2+4b^2Z^2Y^2\right\}\\
 &+\frac{36}{b^4}\left[\left(Z+\frac2b\right)^2+b^2Y^2\right]+\frac{36}{b^6}.
\end{aligned}
\end{equation*}
The auxiliary function for the lump solution in \eqref{four-lump} is
\begin{equation*}
\begin{aligned}
f_{6,8}=3025\Bigl\{&Y^{8} b^{8}+4 Y^{6} b^{4}\Bigl[Z^{2} b^{2} - 4 Z b - 12 b^{3} t + 10\Bigr]\\
&{}+2 Y^{4}\Bigl[3 Z^{4} b^{4} - 16 Z^{3} b^{3} + Z^{2} \left(- 24 b^{5} t + 16 b^{2}\right)+Z \left(- 192 b^{4} t - 16 b\right) + 144 b^{6} t^{2} - 96 b^{3} t + 24\Bigr]\\
&{}+\frac{4 Y^{2}}{b^{4}}\Bigl[Z^{6} b^{6} - 4 Z^{5} b^{5} + Z^{4} \left(12 b^{7} t + 18 b^{4}\right)+Z^{3} \left(96 b^{6} t - 16 b^{3}\right) + Z^{2} \left(720 b^{8} t^{2} - 48 b^{5} t + 96 b^{2}\right)\\
&\qquad{}+Z \left(576 b^{7} t^{2} + 384 b^{4} t + 64 b\right) + 1728 b^{9} t^{3} + 864 b^{6} t^{2} + 96\Bigr]\\
&{}+\frac{1}{b^{8}}\Bigl[Z^{8} b^{8} + Z^{6} \left(48 b^{9} t + 16 b^{6}\right) + 32 Z^{5} b^{5}+Z^{4} \left(288 b^{10} t^{2} + 384 b^{7} t + 96 b^{4}\right) + 128 Z^{3} b^{3}\\
&\qquad{}+Z^{2} \left(- 6912 b^{11} t^{3} + 2304 b^{8} t^{2}\right) - 4608 Z b^{7} t^{2}+20736 b^{12} t^{4} + 4608 b^{6} t^{2} + 1536 b^{3} t + 768\Bigr]\Bigr\}.
\end{aligned}
\end{equation*}
The auxiliary function for the lump solution in \eqref{five-lump} is
\begin{equation*}
\begin{aligned}
f_{6,10}=\frac{1764}{25}\Bigl\{&Y^{10} b^{10}+5 Y^{8} b^{6}\Bigl[Z^{2} b^{2} + 13\Bigr]+10 Y^{6} b^{2}\Bigl[Z^{4} b^{4} - 10 Z^{2} b^{2} - 192 Z b^{4} t - 144 b^{6} t^{2} + 40\Bigr]\\
&{}+\frac{10 Y^{4}}{b^{2}}\Bigl[Z^{6} b^{6} - 5 Z^{4} b^{4} + Z^{2} \left(720 b^{8} t^{2} + 60 b^{2}\right)- 240 Z b^{4} t + 2160 b^{6} t^{2} + 120\Bigr]\\
&{}+\frac{5 Y^{2}}{b^{4}}\Bigl[Z^{8} b^{6} + 28 Z^{6} b^{4} + 384 Z^{5} b^{6} t + Z^{4} \left(1440 b^{8} t^{2} + 400 b^{2}\right)\\
&\qquad{}+2880 Z^{3} b^{4} t + Z^{2} \left(14400 b^{6} t^{2} + 1440\right) + 7680 Z b^{2} t+103680 b^{10} t^{4} + 11520 b^{4} t^{2}\Bigr]\\
&{}+\frac{1}{b^{10}}\Bigl[Z^{10} b^{10} + 25 Z^{8} b^{8} + Z^{6} \left(- 1440 b^{12} t^{2} + 200 b^{6}\right)- 2400 Z^{5} b^{8} t - 7200 Z^{4} b^{10} t^{2} - 9600 Z^{3} b^{6} t\\
&\qquad{}+Z^{2} \left(518400 b^{14} t^{4} + 86400 b^{8} t^{2}\right) + 345600 Z b^{10} t^{3}+518400 b^{12} t^{4} + 115200 b^{6} t^{2} + 14400\Bigr]\Bigr\}.
\end{aligned}
\end{equation*}
The auxiliary function for the lump solution in \eqref{33lump} is
\begin{equation*}
\begin{aligned}
f_{6,12}={}&518400 T^{4}+T^{3}\Bigl[259200 Y^{2} Z b^{2} - 86400 Z^{3}\Bigr]\\
&{}+\frac{T^{2}}{b^{6}}\Bigl[- 8640 H Z b^{6} - 10800 Y^{4} Z^{2} b^{10} + 32400 Y^{2} Z^{4} b^{8} + 64800 Y^{2} Z^{2} b^{6} + 2160 Z^{6} b^{6}\\
&\qquad{}+5040 Y^{6} b^{12} + 32400 Y^{4} b^{8} - 64800 Y^{2} b^{4} + 32400 Z^{4} b^{4} + 194400 Z^{2} b^{2} + 453600\Bigr]\\
&{}+\frac{T}{b^{8}}\Bigl[- 360 Y^{8} Z b^{16} - 960 Y^{6} Z^{3} b^{14} - 6480 Y^{6} Z b^{12} - 720 Y^{4} Z^{5} b^{12}\\
&\qquad- 10800 Y^{4} Z^{3} b^{10} - 16200 Y^{4} Z b^{8} - 2160 Y^{2} Z^{5} b^{8} - 75600 Y^{2} Z^{3} b^{6}\\
&\qquad- 720 H Y^{4} b^{12} - 2160 H Y^{2} b^{8} + 720 H Z^{4} b^{8} + 2160 H Z^{2} b^{6} + 48600 Y^{2} Z b^{4} + 120 Z^{9} b^{8}\\
&\qquad{}+4320 H b^{4} + 2160 Z^{7} b^{6} + 9720 Z^{5} b^{4} - 37800 Z^{3} b^{2} + 194400 Z\Bigr]\\
&{}+\frac{1}{b^{12}}\Bigl[60 H Y^{6} Z b^{18} - 60 H Y^{4} Z^{3} b^{16} + 360 H Y^{4} Z b^{14} - 108 H Y^{2} Z^{5} b^{14}\\
&\qquad- 720 H Y^{2} Z^{3} b^{12} + 540 H Y^{2} Z b^{10} + 6 Y^{10} Z^{2} b^{22} + 15 Y^{8} Z^{4} b^{20} + 180 Y^{8} Z^{2} b^{18}\\
&\qquad{}+20 Y^{6} Z^{6} b^{18} + 360 Y^{6} Z^{4} b^{16} + 2160 Y^{6} Z^{2} b^{14} + 15 Y^{4} Z^{8} b^{16} + 360 Y^{4} Z^{6} b^{14}\\
&\qquad{}+2700 Y^{4} Z^{4} b^{12} + 9450 Y^{4} Z^{2} b^{10} + 6 Y^{2} Z^{10} b^{14} + 180 Y^{2} Z^{8} b^{12}\notag\\
&\qquad{}+36 H^{2} Y^{2} b^{14} + 36 H^{2} Z^{2} b^{12} + 1440 Y^{2} Z^{6} b^{10} + 4050 Y^{2} Z^{4} b^{8} + 28350 Y^{2} Z^{2} b^{6}\\
&\qquad{}+12 H Z^{7} b^{12} + 72 H Z^{5} b^{10} - 180 H Z^{3} b^{8} + 540 H Z b^{6} + Z^{12} b^{12} + 36 Z^{10} b^{10} + 270 Z^{8} b^{8}\\
&\qquad{}+Y^{12} b^{24} + 36 Y^{10} b^{20} + 630 Y^{8} b^{16} + 990 Z^{6} b^{6} - 6075 Z^{4} b^{4} + 60750 Z^{2} b^{2}\\
&\qquad{}+36 H^{2} b^{10} + 6390 Y^{6} b^{12} + 34425 Y^{4} b^{8} + 93150 Y^{2} b^{4} + 50625\Bigr].
\end{aligned}
\end{equation*}

\end{sloppypar}
\end{document}